\documentclass[11pt]{article}
\usepackage{xcolor}
\usepackage{cite}
\usepackage{multirow}
\usepackage{enumerate}
\usepackage[scale=0.8]{geometry}
\usepackage{subfigure}
\usepackage{amssymb}
\usepackage{amsfonts,amsmath}
\usepackage{multirow,multicol}
\usepackage{bbm}
\usepackage{xspace}
\usepackage{slashed}
\usepackage{graphicx}

\usepackage{hyperref}

\begin{document}
\begin{center}
{\huge\bfseries Level Crossings, Attractor Points and Complex Multiplication}\\[10mm]

Hamza Ahmed$^{a,}$\footnote{\href{mailto:ahmed.ha@northeastern.edu}{ahmed.ha@northeastern.edu}}, Fabian Ruehle$^{a,b,c,}$\footnote{\href{mailto:f.ruehle@northeastern.edu}{f.ruehle@northeastern.edu}}\\[10mm]
\bigskip
{
	{\it ${}^{\text{a}}$ Department of Physics, Northeastern University, Boston, MA 02115, USA}\\[.5em]
	{\it ${}^{\text{b}}$ Department of Mathematics, Northeastern University, Boston, MA 02115, USA}\\[.5em]
	{\it ${}^{\text{c}}$ NSF Institute for Artificial Intelligence and Fundamental Interactions, Boston, MA, USA}\\[.5em]
}
\end{center}
\setcounter{footnote}{0} 
\bigskip\bigskip

\begin{abstract}
We study the complex structure moduli dependence of the scalar Laplacian eigenmodes for one-parameter families of Calabi-Yau $n$-folds in $\mathbbm{P}^{n+1}$. It was previously observed that some eigenmodes get lighter while others get heavier as a function of these moduli, which leads to eigenvalue crossing. We identify the cause for this behavior for the torus. We then show that at points in a sublocus of complex structure moduli space where Laplacian eigenmodes cross, the torus has complex multiplication. We speculate that the generalization to arbitrary Calabi-Yau manifolds could be that level crossing is related to rank one attractor points. To test this, we compute the eigenmodes numerically for the quartic K3 and the quintic threefold, and match crossings to CM and attractor points in these varieties. To quantify the error of our numerical methods, we also study the dependence of the numerical spectrum on the quality of the Calabi-Yau metric approximation, the number of points sampled from the Calabi-Yau variety, the truncation of the eigenbasis, and the the distance from degeneration points in complex structure moduli space.
\end{abstract}

\clearpage
\tableofcontents

%%%%%%%%%%%%%%%%%%%%%%%%%%%%%%%%%%%%%%%%%%%%%%%%%%%%%%%%%%%%%%%%%%%%%%%%%%%%
\section{Introduction}
String theory has a history of bringing together many different fields within Physics and Mathematics. In the context of String Geometry or String Phenomenology, one often studies string compactifications on Calabi-Yau (CY) $n$-folds to obtain lower-dimensional theories. Candelas et.al.~showed~\cite{Candelas:1985en} that such reductions can preserve supersymmetry and hence computational control, due to the existence of a Ricci-flat metric on these spaces~\cite{Calabi:1957aaa, Yau:1977ms}. In the absence of an analytic expression for the metric, properties of the lower-dimensional theory were derived using powerful tools from algebraic geometry. This lead to the discovery of new mathematics, such as mirror symmetry~\cite{Candelas:1990rm}.

However, algebraic geometry is not the only mathematical field with connections to string theory. In his seminal work~\cite{Moore:1998pn}, Moore established a connection between Calabi-Yau compactifications and number theory.  A few years prior to Moore, Ferrara et.al.~\cite{Ferrara:1995ih} introduced attractor varieties. The authors observed that dyonic BPS black holes in 4D $\mathcal{N}=2$ theories, the vector multiplets of the theory flow to fixed points called attractors. In the context of string theory, upon compactifying Type IIB theory on a CY threefold, the vector multiplet moduli space is mapped to the complex structure moduli space, and the attractor flow becomes a flow to fixed points in this moduli space. Moore made the remarkable observation that, in the case where the CY is $T^6$ or $K3\times T^2$, these attractor points correspond to special points where the torus has a number-theoretic property called complex multiplication (CM). In these cases, attractors are arithmetic.

A lot of research has gone into attractor points and complex multiplication since. Shortly after Moore, Gukov and Vafa observed that there is a potential tension between Rational Conformal Field theories being dense in the set of all superconformal field theories and the scarcity of CM points, based on a number theory conjecture due to Andr\'e~\cite{Andre:1989aaa} and Oort~\cite{Oort:1997aaa}, which has been proven in full generality very recently~\cite{Pila:2022aaa}. Also very recently, a set of counterexamples to one of Moore's conjectures has been constructed for (complex odd-dimensional) CYs of sufficiently high dimension~\cite{Lam:2020qge}. However, Moore's conjecture seems to be true for rank 2 attractor points, where it is ``protected'' by the Hodge conjecture~\cite{Hodge:1950aaa}, as recently explored by Candelas et.al.~\cite{Candelas:2019llw}.

With the advent of powerful machine learning techniques and their applications in string theory~\cite{He:2017set,Krefl:2017yox,Ruehle:2017mzq,Carifio:2017bov}, it became possible to perform complex numerical computations fast and with high accuracy~\cite{Ruehle:2020jrk}. Building on earlier work~\cite{Donaldson:2005aaa, Douglas:2006rr, Braun:2008jp}, neural networks were developed to approximate the Ricci-flat metric on CY manifolds~\cite{Ashmore:2019wzb,Anderson:2020hux,Douglas:2020hpv,Jejjala:2020wcc,Ashmore:2021ohf,Larfors:2021pbb,Larfors:2022nep,Gerdes:2022nzr}. One exciting application which requires knowledge of the Ricci-flat metric is computing the spectrum of the compactified theory: while the zero modes can be counted by computing cohomologies due to Hodge theory, there is currently no known way to obtain the massive spectrum. As a consequence, not much is known about the full eigenspectrum of Laplacian operators on Calabi-Yau manifolds. On the other hand, the swampland distance conjecture (SDC)~\cite{Ooguri:2006in}, which has received a lot of attention recently, states that the mass gap of the eigenspectrum closes exponentially fast as a function of the geodesic distance between two points in moduli space. In~\cite{Ashmore:2021qdf}, Ashmore and one of the authors studied the spectrum of the scalar Laplacian on (a one-parameter family of) quintic CY threefolds with the goal to check -- and verify in that case -- the swampland distance conjecture. 

The one-parameter family of quintics has only one complex structure parameter instead of 101, which means that it has a large symmetry group. The Laplacian eigenmodes arrange themselves in irreducible representations (irreps) of this symmetry group~\cite{Braun:2008jp,Ashmore:2021qdf}. As a consequence, the spectrum is degenerate in codimension 0 with multiplicities given by the dimension of the irreps. However, we observed~\cite{Ashmore:2021qdf} that some eigenmodes became heavier and other became lighter as a function of the geodesic distance along a trajectory of real codimension 1 within the complex structure moduli space of real dimension 2. At points in complex structure moduli space where two modes cross, the number of degenerate eigenmodes does in general not match the dimension of the irreps. Albeit not in contradiction with no-crossing theorems of quantum mechanics (since they appear in codimension 1) or random matrix theory (the scalar Laplacian is a hermitian matrix, but not random and the trajectory is not a simple homotopy due to the complex structure moduli dependence of the metric), we found the observation still surprising.

But even besides the aforementioned points, the observation of crossings begs the questions which we will address in this paper:
\begin{enumerate}
\item What governs the behavior of the eigenmodes, i.e., why do some modes become lighter while others become heavier?
\item What is special about the points in complex structure moduli space where eigenvalue crossings occur? 
\end{enumerate}

The rest of this paper is organized as follows: In Section~\ref{sec:Cubic}, we study the one-parameter family of cubics in $\mathbbm{P}^2$ analogous to the quintic family. This family describes tori, for which we have full analytic control over the metric, the eigenmodes etc. We work out the map between the ambient space description of the torus embedded as a hypersurface in $\mathbbm{P}^2$ and a description of the ``flat'' tours on a lattice $\Lambda$ and make the observation that crossing points are related to special points in complex structure moduli space where the torus has complex multiplication. In Section~\ref{sec:XingAndCM}, we review how to extend the notion of CM to arbitrary CY manifolds by phrasing it as a condition on their middle cohomology. As we will review, the one-parameter family of K3 surfaces admits CM points where its Picard rank jumps from 19 to to 20, while the only known CM point for the one-parameter family of quintics is at the Fermat point. We then review the relation between CM points and attractor points and argue that it is more likely that crossings are related to attractor points. In Section~\ref{sec:NumericSpectrum}, we turn to the numerical analysis part of the paper. We describe how we approximate the Ricci flat metric and the Laplacian eigenspectrum. We study the different approximation steps and their influence on the accuracy by comparing the numerical and the analytic result on $T^2$. We do so in order to get an idea on the errors to expect when checking the relation between attractor points and eigenvalue crossings in Section~\ref{sec:NumericAnalysis} (but this might be of independent interest in future applications involving the Laplacian eigenspectrum). We end with conclusions and an outlook in Section~\ref{sec:Conclusions}.

%%%%%%%%%%%%%%%%%%%%%%%%%%%%%%%%%%%%%%%%%%%%%%%%%%%%%%%%%%%%%%%%%%%%%%%%%%%%
\section{Fermat cubic}
\label{sec:Cubic}
We start our analysis with the one-parameter family of Fermat cubics in $\mathbbm{P}^2$. These correspond to Calabi-Yau one-folds, i.e., tori. These simple cases allow for an exact treatment of the metric, the spectrum and the eigenfunctions. We work out the map from the description of a $T^2$ embedded as a hypersurface in $\mathbbm{P}^2$ with one complex structure parameter, to the flat torus described by one complex coordinate $w\in\mathbbm{C}/\Lambda$ on a lattice $\Lambda$ with periods $1$ and $\tau\in\mathbbm{C}$.\footnote{We may assume without loss of generality $\tau\in\mathbbm{H}=\{\tau\in\mathbbm{C}~|~\text{Im}(\tau)\geq0\}$.}

The one-parameter family of Fermat cubics in $\mathbbm{P}^2$ is given by
\begin{align}
\label{eq:cubic}
X=z_{0}^{3}+z_{1}^{3}+z_{2}^{3}-3\psi z_{0}z_{1}z_{2}=0
\end{align}
where the $z_{i}$ are the homogeneous coordinates of $\mathbbm{P}^2$, and $\psi\in \mathbbm{C}$ is a single complex structure modulus related to $\tau$. The manifold is singular at $\psi=1$ and $|\psi|\to\infty$. The latter is the large complex structure degeneration where the manifold degenerates into three complex lines, and the former is the analog of the conifold point in the quintic, where $X=dX=0$ (e.g.\ $[z_0:z_1:z_2]=[1:1:1]$). However, unlike the conifold on the quintic, this point is dual to the large complex structure point. 

Since the hypersurface is Calabi-Yau, it admits a Ricci-flat metric (which, in the case of $d=1$, is actually flat). We also have a nowhere vanishing holomorphic 1-form $\Omega$, which we can integrate over the two 1-cycles $A$ and $B$ to obtain the two periods of the torus lattice $\Lambda$. Since we want to choose a basis of 1 and $\tau$, we divide by the period of the $A$-cycle,
\begin{align}
\label{eq:tau}
\tau=\frac{\int_{B} \Omega_{1} }{\int_{A} \Omega_{1} }
\end{align}
The periods can be obtained from the underlying Picard Fuchs system using standard methods~\cite{Candelas:1990rm,Hosono:1993qy,Hosono:1994ax}. The solutions are given in terms of hypergeometric functions
\begin{align}
\label{eq:tauHGF}
\tau=\tau(\psi)=\frac{i}{\sqrt{3}}\;\frac{{}_2F_1\left(\frac{1}{3},\frac{2}{3};1;1-\frac{1}{\psi^{3}}\right)}{{}_2F_1\left(\frac{1}{3},\frac{2}{3};1;\frac{1}{\psi^{3}}\right)}\,.
\end{align}

Next, we map the cubic into Weierstrass form
\begin{align}
\label{eq:weierstrassform}
y^{2}z=4x^{3}-g_{2}(\psi)xz^{2}-g_{3}(\psi)z^{3} \,.
\end{align}
We find that the curve is smooth in the patch $z=1$, so we will work in this patch from now on. Transforming~\eqref{eq:cubic} into~\eqref{eq:weierstrassform} can be done in general using Nagell's algorithm~\cite{Nagell:1928aaa}. However, the symmetries of the Fermat cubic make the problem very simple, leading for example to the transformation
\begin{align}
\label{weimap}
\left(\begin{array}{ccc}
x\\
y \\
z \\
\end{array}\right)
=M^{-1}
\left(\begin{array}{ccc}
z_{0}\\
z_{1}\\
z_{2} \\
\end{array}\right)\,, 
\qquad
M=
\left(\begin{array}{ccc}
-\frac{2^{2 / 3}}{\left(1-\psi^{3}\right)^{1 / 3}} & 0 & -\frac{\psi^{2}}{4\left(1-\psi^{3}\right)} \\
\frac{\psi}{2^{1 / 3}\left(1-\psi^{3}\right)^{1 / 3}} & -1 & \frac{4-\psi^{3}}{24\left(1-\psi^{3}\right)} \\
\frac{\psi}{2^{1 / 3}\left(1-\psi^{3}\right)^{1 / 3}} & 1 & \frac{4-\psi^{3}}{24\left(1-\psi^{3}\right)}
\end{array}\right)\,.
\end{align}
The solution of the Weierstrass equation~\eqref{eq:weierstrassform} in terms of the flat torus coordinate is the Weierstrass $\wp(w)$ function and its derivative,
\begin{align}
\label{weimap2}
x=\wp(w)\,,\qquad y=\wp'(w)\,, \qquad z=1\,.
\end{align}

Next, we read off $g_2(\psi)$ and $g_3(\psi)$,
\begin{align}
\label{eq:EisensteinFcts}
g_{2}(\psi)=\frac{\psi\left(8+\psi^{3}\right)}{24 \times 2^{\frac13}\left(1-\psi^{3}\right)^{\frac43}}\,,\qquad\qquad
g_{3}(\psi)=\frac{8+20 \psi^{3}-\psi^{6}}{864\left(1-\psi^{3}\right)^{2}}\,.
\end{align}
To find the complex structure parameter $\tau$ as a function of $\psi$, we can either use~\eqref{eq:tauHGF}, or express $\tau$ in terms of $\psi$ via the Klein $j$ function
\begin{align}
\label{eq:Kleinj}
j(\tau)=\frac{g_{2}(\psi)^{3}}{g_{2}(\psi)^{3}-27g_{3}(\psi)^{2}}
\end{align}
In either case, an analytic expression only exists for very special values of $\psi$, so we will have to resort to numerical formulae to obtain $\tau$ as a general function of $\psi$. However, we find that the special values $\tau\in\{i,e^{2\pi i/3}, i\infty\}$ are obtained for
\begin{align}
\label{eq:SpecialValuesTau}
\begin{array}{rcrl}
\tau=i&\Leftrightarrow&g_{3}=0\qquad\Leftrightarrow& \qquad \psi=(1\pm\sqrt{3})\xi^k\\\
\tau=e^{2\pi i/3}&\Leftrightarrow&g_{2}=0\qquad\Leftrightarrow&\qquad \psi=0\quad\text{or}\quad \psi=-2\xi^k\\
\tau=i\infty&\Leftrightarrow& g_2^3-27g_{3}^2=0\qquad\Leftrightarrow&\qquad \psi=\xi^k\quad\text{or}\quad |\psi|\to\infty
\end{array}
\end{align}
where $k=0,1,2$ and $\xi=e^{2\pi i/3}$ is a third root of unity. Note that only $\psi^3$ enters in the hypergeometric functions~\eqref{eq:tauHGF}. Alternatively, note that $\psi\to\xi\psi$ in~\eqref{eq:cubic} can be undone by $x_0\to\xi^{-1} x_0$, say. This means that a good coordinate on the complex structure moduli space is $\psi^3$ rather than $\psi$. More generally, we should use $\psi^{n+2}$ for the one-parameter family of CY $n$-folds discussed in this paper.

\begin{figure}[t]
\centering
\includegraphics[width=.34\textwidth]{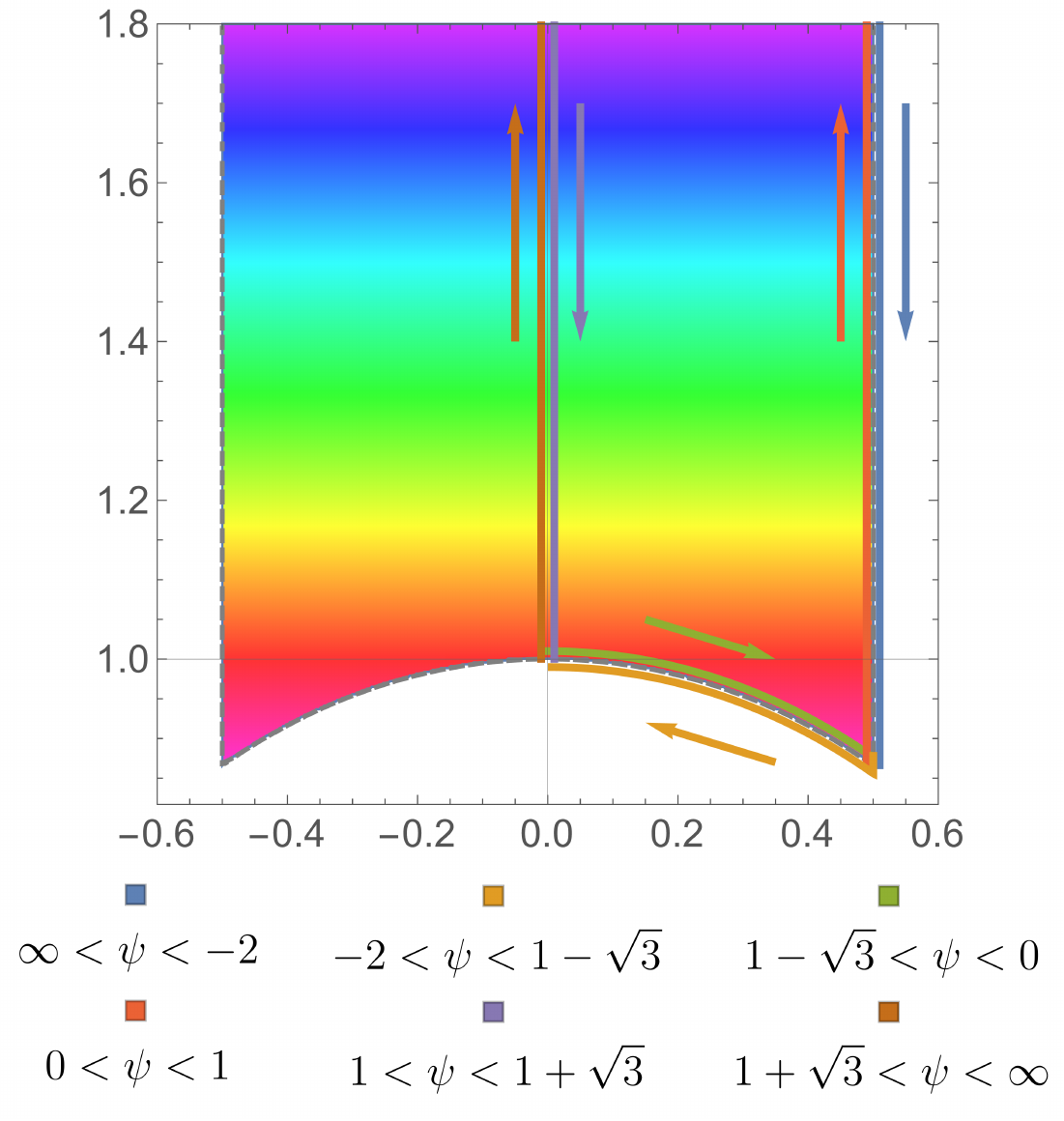}\qquad\qquad
\includegraphics[width=.45\textwidth]{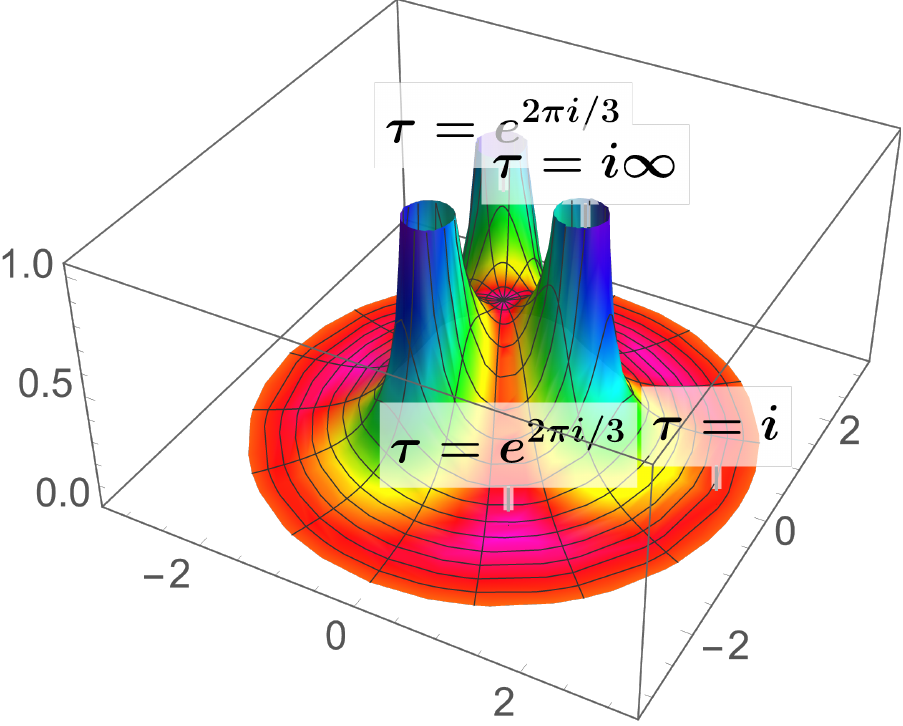}
\caption{Left: A fundamental domain for $\tau$ in the upper half plane is plotted with dashed gray lines. The way $\tau(\psi)$ varies as $\psi\in\mathbbm{R}$ is varied is depicted with arrows; note that $\tau$ traverses the boundary of the fundamental domain as well as the line $\text{Re}(\tau)=0$ twice (and in opposite directions). Colors in the interior correspond to the value of the metric. Right: Moduli space metric with coordinate $\psi$. Note that the fundamental domain is only a third of this.}
\label{fig:ModuliSpaceMetric}
\end{figure}

The modular parameter $\tau$ can be defined in the complex upper half-plane, modulo $\text{SL}(2,\mathbbm{Z})$ transformations. Note that for $\psi$ real, $\tau$ only takes values along (half) the boundary of the fundamental domain and along $\text{Re}(\tau)=0$. The different regions of moduli space traced out for these values, together with the whole fundamental domain for $\tau$, are shown on the left of Figure~\ref{fig:ModuliSpaceMetric}. The color coding corresponds to the values of the moduli space metric. In terms of $\tau$, the Weil-Petersson metric on the Teichm\"uller space is
\begin{align}
\label{eq:metric}
g_{\tau\bar\tau}=\frac{1}{\text{Im}(\tau)^2}\,,
\end{align}
so the metric is constant along constant slices of $\text{Im}(\tau)$. In terms of $\psi$, the moduli space metric can be obtained from the periods~\cite{Candelas:1990rm}, or by inserting $\tau(\psi)$ in~\eqref{eq:metric}. The resulting metric $g_{\psi\bar\psi}$ is plotted on the right of Figure~\ref{fig:ModuliSpaceMetric}. The color coding matches the color coding for the metric $g_{\tau\bar\tau}$.

\subsection{Discrete symmetries}
\label{sec:DiscreteSymmetries}
All hypersurfaces \eqref{eq:cubic} have an $S_{3}$ symmetry which arises from invariance under permutation of the $z_{i}$, and a $\mathbbm{Z}_{3}$ symmetry which arises from invariance under multiplying the $z_{i}$ by a phase. Naively, we have three $\mathbbm{Z}_{3}$ factors acting as
\begin{align}
&(z_{0},z_{1},z_{2}) \xrightarrow{\mathbbm{Z}_{3}^{(1)}} (~~~\xi z_{0},~\xi^{-1}z_{1},~\phantom{\xi^{-1}}z_{2})\,, \nonumber \\
&(z_{0},z_{1},z_{2}) \xrightarrow{\mathbbm{Z}_{3}^{(2)}} (\phantom{\xi^{-1}} z_{0},~~~~\xi z_{1},~\xi^{-1}z_{2})\,,\\
&(z_{0},z_{1},z_{2}) \xrightarrow{\mathbbm{Z}_{3}^{(3)}} (\xi^{-1}z_{0},~\phantom{\xi^{-1}}z_{1},~~~~\xi z_{2})\,. \nonumber
\end{align}
where $\xi=e^{2\pi i/3}$. However, only one of these is independent, since the others are related via projective scalings. We choose without loss of generality $\mathbbm{Z}_{3}^{(1)}$ and call the corresponding generator $p$. 

Along more special points in complex structure moduli space, this symmetry is enhanced: If $\psi\in\mathbbm{R}$, there is a $\mathbbm{Z}_{2}$ symmetry corresponding to complex conjugation. We denote the generator of this symmetry operation by $c$. At the point $\psi=0$, the $\mathbbm{Z}_3$ symmetry is enhanced to $\mathbbm{Z}_3\times\mathbbm{Z}_3$, where we can choose any two of the following three symmetries (the third is not independent but related to the other two by the projective scaling of $\mathbbm{P}^2$):
\begin{align}
&(z_{0},z_{1},z_{2}) \xrightarrow{\mathbbm{Z}_{3}'} (\xi z_{0},z_{1},z_{2})\,, \nonumber \\
&(z_{0},z_{1},z_{2}) \xrightarrow{\mathbbm{Z}_{3}''} (z_{0},\xi z_{1},z_{2})\,,\\
&(z_{0},z_{1},z_{2}) \xrightarrow{\mathbbm{Z}_{3}'''} (z_{0},z_{1},\xi z_{2})\,. \nonumber
\end{align}
These symmetries are key in our analysis, since the eigenvalues of the scalar Laplacian occur in irreducible representations (irreps) of the combined symmetry group~\cite{Braun:2008jp}. We start our discussion with $S_3$, which is generated by cyclic permutations ($s$) and transpositions ($t$). These act on the homogeneous coordinates $[z_{0}:z_{1}:z_{2}]$ as
\begin{align}
(z_{0},z_{1},z_{2}) \xrightarrow{s} (z_{1},z_{2},z_{0})\,,\qquad (z_{0},z_{1},z_{2}) \xrightarrow{t} (z_{1},z_{0},z_{2})\,.
\end{align}

\begin{table}[t]
\centering
\begin{tabular}{|c||cc|ccc|cccc|}
\hline
$\psi$&\multicolumn{2}{c|}{$\psi\in\mathbbm{C}-\mathbbm{R}$}&\multicolumn{3}{c|}{$\psi\in\mathbbm{R}$}&\multicolumn{4}{c|}{$\psi=0$}\\
\hline
Symmetry&\multicolumn{2}{c|}{$G=S_{3} \rtimes \mathbbm{Z}_{3}$}&\multicolumn{3}{c|}{$(S_{3} \times \mathbbm{Z}_{2}) \rtimes \mathbbm{Z}_{3}$}&\multicolumn{4}{c|}{$(S_{3} \times \mathbbm{Z}_{2}) \rtimes (\mathbbm{Z}_{3}\times\mathbbm{Z}_3)$}\\
\hline
dim(irrep)		&~~~1&2&	~~1&~~2&4	&~~1&~~2&~~4&6\\
\hline
number(irreps)	&~~~2&4&	~~4&~~4&1	&~~4&~~4&~~1&2\\
\hline
\end{tabular}
\caption{Symmetry group of the cubic for different values of the complex structure parameters $\psi$.}
\label{tab:SymmetriesT2}
\end{table}

Note that permutations and complex conjugation commute, while complex conjugation and the $\mathbbm{Z}_3$ action(s) do not. For this reason, the full symmetry group is a semi-direct product of these symmetry operations, cf.~Table~\ref{tab:SymmetriesT2}. The combined symmetry group is
\begin{align}
\psi\in\mathbbm{C}-\mathbbm{R}:&\qquad S_{3} \rtimes \mathbbm{Z}_{3}\,,\nonumber\\
\psi\in\mathbbm{R}:&\qquad (S_{3} \times \mathbbm{Z}_{2}) \rtimes \mathbbm{Z}_{3}\,,\\
\psi=0:&\qquad (S_{3} \times \mathbbm{Z}_{2}) \rtimes (\mathbbm{Z}_{3}\times\mathbbm{Z}_3)\,.\nonumber
\label{cubirrep}
\end{align}
To fully specify the semi-direct product, we need to specify the twisting, i.e., how elements $e \in \mathbbm{Z}_{3}$ change under $g^{-1} \circ e \circ g$, where $g$ is an element of $S_{3}$ (or $S_3 \times \mathbbm{Z}_{2}$). More specifically, we have to specify the action of the generators $s$, $t$ and $c$ on the generator $p$ of $\mathbbm{Z}_{3}$ modulo projective scalings. For $c$, we have the simple action (because $\xi \rightarrow \xi ^{-1}$ under complex conjugation):
\begin{align}
p\xrightarrow{c^{-1}\; \circ\; e\; \circ\; c\;} p^{-1}
\end{align}
For the generators $s$ and $t$, we need to compute their action on any element $e \in \mathbbm{Z}_{3}$. A general action of $\mathbbm{Z}_{3}$ is given by $p^{n}$, where $n$ is an integer (mod 3), by
\begin{align}
(z_{0},z_{1},z_{2}) \rightarrow (\xi ^{n} z_{0},\xi ^{-n} z_{1},z_{2}) \,.
\end{align}
This implies an action on the generators given by
\begin{align}
p\xrightarrow{t^{-1}\; \circ\; e\; \circ\; t\;} p^{-1}\,,\qquad p\xrightarrow{s^{-1\;} \circ\; e\; \circ\; s\;} p
\end{align}
We work out the dimensions $d$ of the irreps and the number $n_d$ of often they occur with the software GAP in Sage~\cite{sagemath}. The results are summarized in Table~\ref{tab:SymmetriesT2}.

\subsection{Eigenvalue analysis}
\label{sec:EigValTorus}
\begin{figure}[t]
\centering
\includegraphics[width=.5\textwidth]{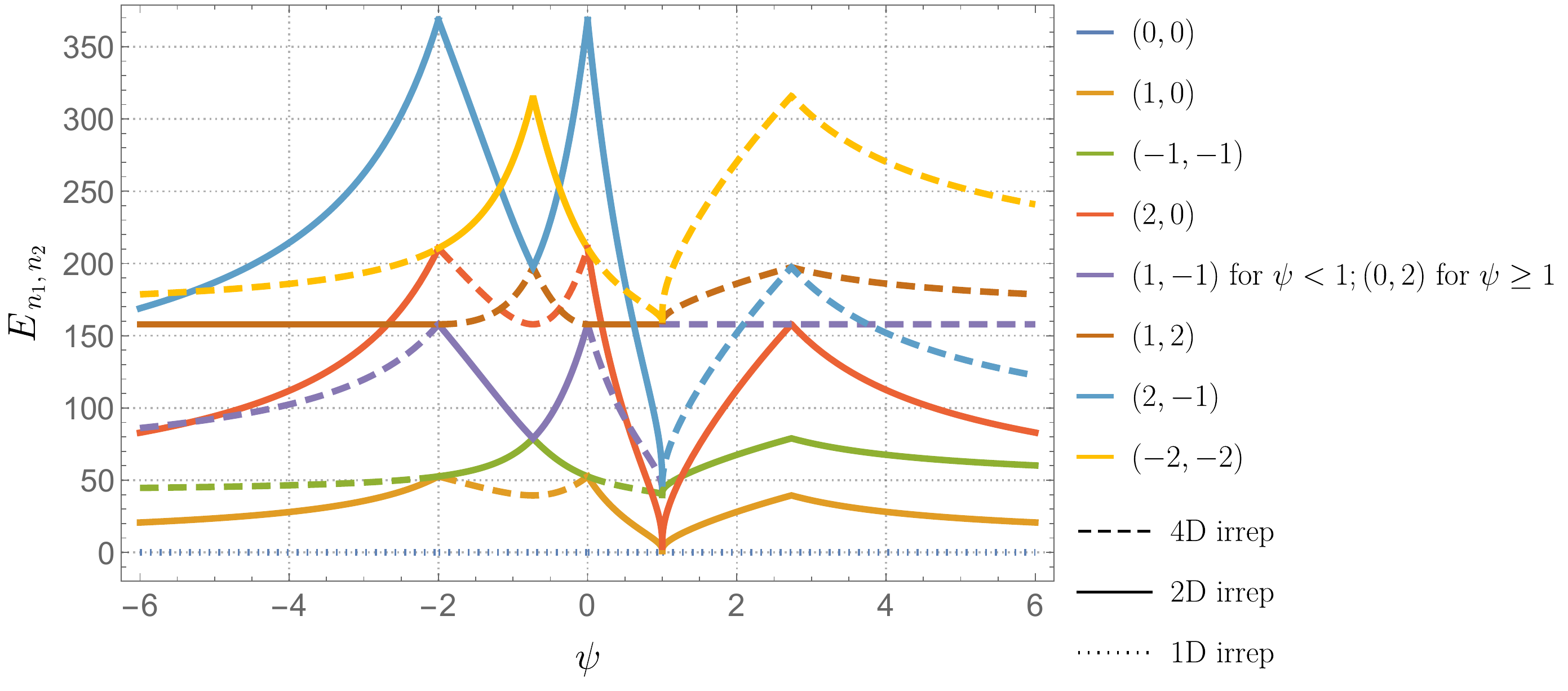}
\includegraphics[width=.42\textwidth]{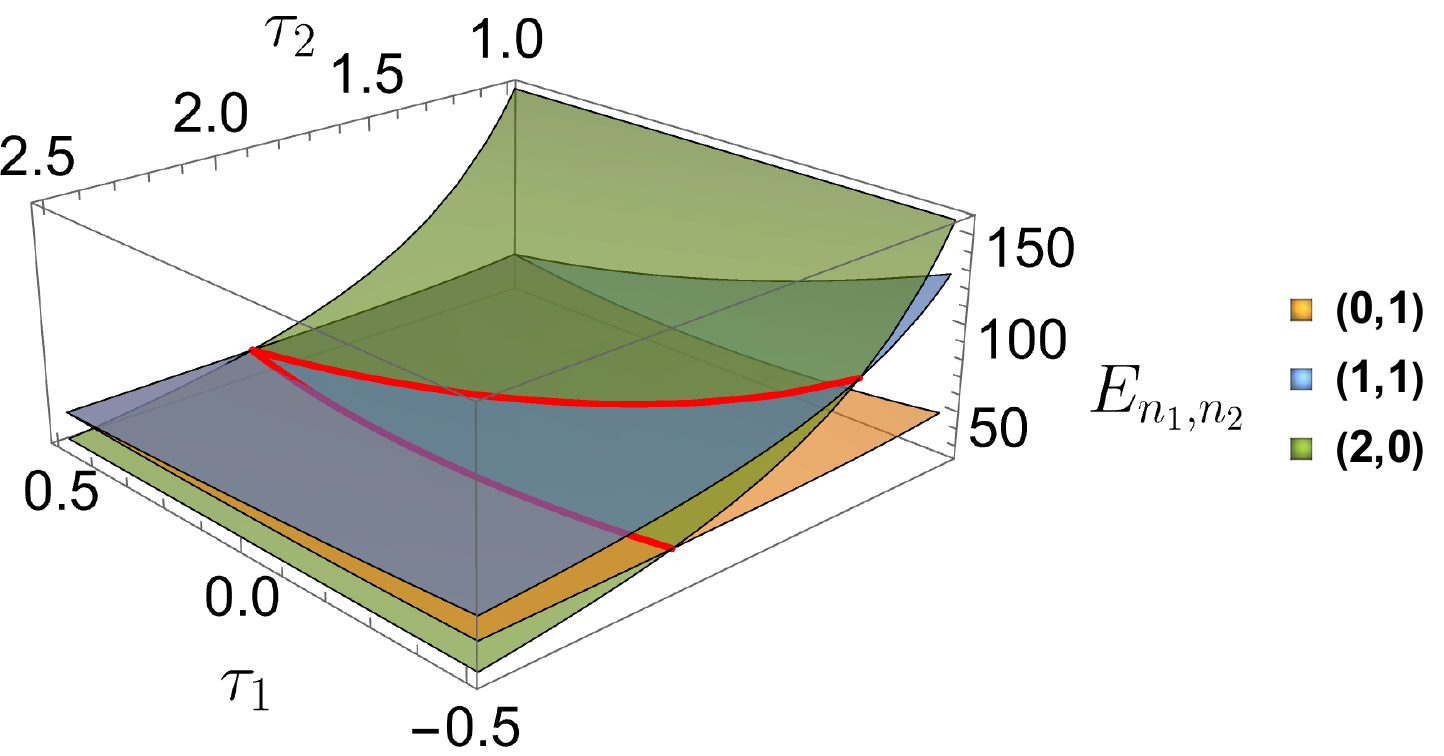}
\caption{Left: The exact spectrum of eigenvalues for the first few modes of the Laplacian on the torus, as a function of the complex structure $\psi$. The different modes correspond to different colors, and the dimension of the irrep corresponds to the type of line. The non-smooth behavior of the spectrum as a function of $\psi$ is explained by the corresponding path through moduli space in terms of $\tau(\psi)$ and monodromies around the infinite distance point at $\psi=\xi^k$ of $\tau(\psi)$ as shown in Figure~\ref{fig:ModuliSpaceMetric}. Right: Eigenmode crossings as $\tau\in\mathbbm{C}$ is varied.}
\label{fig:cross}
\end{figure}

We recall the eigenvalues and eigenfunctions of the scalar Laplacian on the torus in terms of the coordinate $w$. The eigenvalues are given by (setting the K\"ahler modulus to 1)~\cite{Ghilencea:2005vm}:
\begin{align}
\label{eq:eigenvalues}
E_{n_{1},n_{2}}=\frac{4\pi^{2}}{A\tau_{2}}|n_{1}-n_{2}\tau|^{2}=\frac{4\pi^{2}}{\tau_{2}^2}|n_{1}-n_{2}\tau|^{2}\,,
\end{align}
where $n_{1}$ and $n_{2}$ are integers parameterizing the windings around the two torus cycles with periods $1$ and $\tau=\tau_{1}+i \tau_{2}$, respectively, and $A=1\times\tau_2$ is the area of the torus. The corresponding eigenfunctions are
\begin{align}
\label{eq:eigenvectors}
F_{n_1,n_2}(w,\bar{w})=\frac{1}{\sqrt{A}}e^{2\pi i(cw+\bar{c}\bar{w})}=\frac{1}{\sqrt{\tau_{2}}}e^{2\pi i(cw+\bar{c}\bar{w})}\,,\qquad c=\frac{1}{2\tau_{2}}\left(n_{1}(1+i\tau_{1})-i n_{2}\right)\,.
\end{align}
We plot the eigenvalues $E_{n_{1},n_{2}}$ as a function of $\psi\in\mathbbm{R}$, using the implicit map~\eqref{eq:Kleinj}, in Figure~\ref{fig:cross} on the left. More generally, the eigenmodes vary as a function of $\tau$ or $\psi\in\mathbbm{C}$ and intersect along a real codimension 1 line. This crossing line is illustrated for three modes in Figure~\ref{fig:cross} on the right.

We see that the cubic spectrum shows a qualitatively similar behavior to the one observed for the quintic in~\cite{Ashmore:2021qdf}: It was already known that the multiplicity of coincident eigenvalues in codimension 0 in complex structure moduli space is in accordance with the dimensions of the irreps of the symmetry group of the manifold~\cite{Braun:2008jp}. Moreover, it was observed in~\cite{Ashmore:2021qdf} that some irreps become heavier while others become lighter and hence they cross as one moves in complex structure moduli space. At the crossings, the symmetry group does not enhance (except at $\psi=0$), and hence the multiplicities along the real codimension 1 locus where crossings occur do in general not fall into irreps of the symmetry group. We will address the question of what is special about the crossing points in the following sections.

For the torus, we have analytic control over the spectrum. Expanding~\eqref{eq:tauHGF} around large $\psi$, we get
\begin{align}
\label{eq:thyp}
\tau(\psi)=\frac{3 i}{2\pi}\ln(3\psi)\,.
\end{align}
Note that for real $\psi>0$, $\tau$ is purely imaginary, i.e., $\tau_1=0$. Thus, along the red path\footnote{This path is actually a geodesic in complex structure moduli space, as explained in~\cite{Ashmore:2021qdf}} in Figure~\ref{fig:ModuliSpaceMetric}, the expression for the eigenvalues in~\eqref{eq:eigenvalues} simplifies to
\begin{align}
E_{n_{1},n_{2}}=4\pi^{2}\left[\left(\frac{n_{1}}{\tau_{2}}\right)^{2}+n_{2}^{2}\right]\,.
\end{align}

For purely imaginary $\tau$ (or equivalently $\psi\in\mathbbm{R}^+$), there are three different scenarios:
\begin{enumerate}
  \item For $n_{1}=n_{2}=0$, we get the unique zero mode with multiplicity one.
  \item For the cases where either $n_{1}$ or $n_{2}$ vanish, we get:
  \begin{itemize}
      \item $(n_{1}, n_{2})=(0,\pm l)$: the eigenvalues are constant $E_{0,\pm l}=4\pi^2l^{2}$, since the area-suppressed  denominator cancels the numerator, and we keep the other period fixed at length 1.
      \item $(n_{1}, n_{2})=(\pm k, 0)$: the eigenvalues run down quadratically in $\tau_2$, or logarithmically in $\psi$ for $\psi$ large , $E_{\pm k,0}=\frac{4\pi^2k^{2}}{\tau_2^2}\sim\frac{16\pi^4k^{2}}{9\ln(3\psi)^2}$
  \end{itemize}
  These are the two cases of multiplicity two (i.e., transforming in two two-dimensional irreps of the symmetry group); one is constant and one is running down.
  \item For $(n_{1},n_{2})=(\pm k,\pm l)$, all eigenvalues run down logarithmically for large enough $\psi$, following $E_{\pm k,\pm l}=4\pi^2(l^{2}+\frac{k^{2}}{\tau_2^2})\sim\frac{16\pi^4k^{2}}{9\ln(3\psi)^2}$. These energy levels are four-fold degenerate, and hence correspond to the four-dimensional irrep described in Section~\ref{sec:DiscreteSymmetries}.
\end{enumerate}

To summarize, we find that all energy levels in codimension 0 arrange in irreps of the symmetry group. Along a path where $\tau$ is purely imaginary, modes are either constant or run down logarithmically. It is this logarithmic behavior that leads to the exponential dependence of the massive KK tower on the distance in field space as predicted by the swampland distance conjecture~\cite{Ooguri:2006in}. For general $\tau\in\mathbbm{C}$, we observe multiplicities are running up as $\tau_1$ gets smaller and $\tau_2$ is kept fixed, cf.\ Figure~\ref{fig:cross}. In that case, the denominator is constant but $\tau_1$ is in the numerator varies. Note however that $\tau_1\in]-0.5,0.5]$, so the amount by which the modes run up is limited. 

This analysis makes the origin for the running behavior of the modes obvious: We fix one period (the $A$-cycle, say) to 1, while we vary the other (the $B$-cycle) by changing the complex structure parameter $\psi$ (or $\tau$). This means that the eigenmodes associated with the $A$-cycle become lighter as the $B$-cycle volume and with it the volume of the entire CY grows: indeed, the running of the eigenmodes $(\pm k,0)$ is is governed by the overall $1/\text{voulme}$ factor. For modes associated with just wrapping the $B$-cycle, we also get the $1/\text{voulme}$ factor suppression, but at the same time, we get a $|\tau|^2$ factor from the winding. For purely imaginary $\tau$ both factors cancel exactly and the modes are constant; for general $\tau$ the area only depends on $\tau_2$, while $\tau_1$ appears in the numerator and the exact behavior depends on the trajectory in the Teichm\"uller space. Finally, the modes $(\pm k, \pm l)$ that wrap both the $A$- and the $B$-cycle also become lighter as $\tau_2\to\infty$ (or equivalently $\psi\to-\infty$). For a general path through moduli space, either behavior of the mass can occur.

It is very suggestive that a similar effect explains this behavior observed in~\cite{Ashmore:2021qdf} for the quintic as well: There, $\psi$ was varied along a geodesic with $\psi\in\mathbbm{R}$, which connects the conifold point to the large complex structure point. As was worked out by Candelas et.al.~\cite{Candelas:1990rm} for the Fermat quintic, but should hold more generally according to the SYZ conjecture~\cite{Strominger:1996it}, the Fermat quintic has a $T^3$-fibration. At the conifold point, the base $S^3$ shrinks to a point, while in the large complex structure limit, the $T^3$ fiber shrinks. Hence, along the path studied in~\cite{Ashmore:2021qdf}, we have two 3-cycles that compete and can thus lead to some eigenmodes becoming lighter while others become heavier.

\subsection{Level crossing and enhanced symmetries}
Looking at the values of $\psi$ where the crossings occur, there seems in general nothing special in the hypersurface equation~\eqref{eq:cubic} or in the moduli space metric~\eqref{eq:metric} (see also Figure~\ref{fig:ModuliSpaceMetric}): A generic crossing occurs away from the point of maximal unipotent monodromy (MUM), and there are no (apparent) symmetry enhancement of the defining equation at the points where the eigenvalues cross, and hence no explanation for the observed multiplicities at these points. 

However, consider for example the special cases $\psi=-2\xi^k$ and $\psi=0$, corresponding to $\tau=e^{2\pi i/3}$ as shown in~\eqref{eq:SpecialValuesTau}. From Table~\ref{tab:SymmetriesT2}, we see that the latter has an enhanced symmetry by an additional factor of $\mathbbm{Z}_3$, which leads to the existence of a 6D irrep at $\psi=0$. And indeed, we observe crossings of the 4D irreps with the 2D irreps at $\psi=0$ (and consequently also at $\psi=-2$). The two points $\psi=0$ and $\psi=2$ are related by an $\text{SL}(2,\mathbbm{Z})$ transformation on the hypersurface, which is not apparent from the defining equation in $\mathbbm{P}^2$.

While it could be that there is an enhanced, non-linearly realized symmetry at all crossing points, we will now point out another property of the manifold at values of the complex structure where level-crossing occurs. According to~\eqref{eq:eigenvalues}, two modes $(n_{1},n_{2})$ and $(m_{1},m_{2})$ cross when
\begin{align}
\label{taucross}
\tau_1^{2} + \tau_2^{2}=\frac{n_{1}^{2}-m_{1}^{2}}{m_{2}^{2}-n^{2}_{2}} + 2\tau_1\frac{n_{1}n_{2}-m_{1}m_{2}}{m_{2}^{2}-n^{2}_{2}}
\end{align}
These points do not seem special from the point of view of the hypersurface, but the values of $\tau$ in \eqref{taucross} are special: they correspond (for $\tau_1\in\mathbbm{Q}$) to algebraic periods. At these points, the torus has \emph{Complex Multiplication} (CM), which we will explain in more detail in the next section. Moreover, these values correspond to attractor points as first observed by Moore in~\cite{Moore:1998pn}.

Note that the above condition $\tau_1\in\mathbbm{Q}$ is satisfied for $\psi\in\mathbbm{R}$. To see this, we should distinguish the three cases where $\tau$ is purely imaginary, $\tau$ is along the boundary of the fundamental domain with $\tau_1=\frac12$, and $\tau$ is on the boundary circle, cf.\ Figure~\ref{fig:ModuliSpaceMetric}. 

The first case $\tau_1=0$ occurs for $\psi>1$, which means that crossings are at
\begin{align}
\label{taucrossre}
\tau=i\tau_2=i\sqrt{\frac{n_{1}^{2}-m_{1}^{2}}{m_{2}^{2}-n^{2}_{2}}}\,.
\end{align}
For large $\psi$, such that the expansion~\eqref{eq:thyp} is valid, this means that level crossings appear at (we take the branch where $\tau_2>0$)
\begin{align}
\label{psicross}
\psi\approx\frac13\exp\left(\frac{2 \pi }{3} \sqrt{\frac{n_{1}^{2}-m_{1}^{2}}{m_{2}^{2}-n_{2}^{2}}}\right)\,.
\end{align}

Similarly, we have $\tau_1=1/2$ for $\psi\leq-2$ or $0\leq\psi\leq1$, such that level crossings appear at (again, we take the branch where $\tau_2>0$)
\begin{align}
\tau=\frac12+\frac{i}{2}\sqrt{\frac{(2n_{1}-n_{2})^{2}-(2m_1-m_2)^2}{m_{2}^{2}-n^{2}_{2}}}
\end{align}

Finally, for the third case $|\tau|^2=1$, condition~\eqref{taucross} for the crossing becomes
\begin{align}
\tau_1=\frac{(m_1^2 + m_2^2) + (n_1^2 - n_2^2)}{2 (m_1 m_2 - n_1 n_2)}\,.
\end{align}
This means all crossings along the circle occur at $\tau_1\in\mathbbm{Q}$. 

So for all $\psi\in\mathbbm{R}$, CM points occur when two eigenmodes cross. The converse is not true though, i.e., there are more general solutions to~\eqref{taucross}, which appear in (real) codimension 1. Indeed, choosing an arbitrary value for $\tau_1$ (or $\tau_2$) and two eigenmodes, we can just solve~\eqref{taucross} for $\tau_2$ (or $\tau_1$) to find the corresponding complex structure where these eigenmodes cross (if the crossings exist, i.e., if~\eqref{taucross} has a solution with $\tau_1,\tau_2\in\mathbbm{R}$). Nevertheless, these codimension 1 eigenmode crossings can then be followed as in Figure~\eqref{fig:cross} to a point on the boundary or at $\text{Im}(\psi)=0$, corresponding to the cases with rational periods discussed above. 

\section{Attractor points and Complex Multiplication}
\label{sec:XingAndCM}
For tori and K3 surfaces, rank 1 attractors, complex multiplication, and the appearance of rational periods are closely related (for general CYs, this is not the case). In this section, we will discuss these quantities and their relation for the one-parameter families of CY $(n-1)$-folds in $\mathbbm{P}^n$,
\begin{align}
\label{eq:DworkCY}
\sum_{i=0}^{n} z_i^{n+1} - (n+1)\psi \prod_{i=0}^n z_i = 0\,.
\end{align}

\subsection{Complex multiplication}
\subsubsection{The cubic CY 1-fold}
Let us start our discussion by looking at CM tori. We start with a binary quadratic form in two variables with integer coefficients,
\begin{align}
\label{eq:QuadraticForm}
q(x,y)=ax^{2}+bxy+cy^{2}
\end{align}
where $a,b,c \in \mathbbm{Z}$ and $a,c \geq 0$. We will be interested in the case with negative discriminant 
\begin{align}
\label{eq:Discriminant}
D=b^{2}-4ac < 0\,.
\end{align} 

There is an $\text{SL}(2,\mathbbm{Z})$ action defined on these binary quadratic forms, 
\begin{align}
\begin{pmatrix}
x\\y
\end{pmatrix}
\quad\to\quad 
\begin{pmatrix}
p&q\\r&s
\end{pmatrix}
\cdot
\begin{pmatrix}
x\\y
\end{pmatrix}\,,\qquad
\begin{pmatrix}
p&q\\r&s
\end{pmatrix}~\in~\text{SL}(2,\mathbbm{Z})\,.
\end{align}
which amounts to a basis change that changes the coefficients $a,b,c$ but leaves the discriminant invariant. The number of orbits of this action, i.e., the number of equivalence classes of binary quadratic forms up to $\text{SL}(2,\mathbbm{Z})$ transformations, is called the class number and is denoted by $h(D)$. In the case $D<0$, we can find the solutions to~\eqref{eq:QuadraticForm} by extending the rationals by $\sqrt{D}$. This way, we get an \emph{imaginary quadratic field} denoted by $\mathbbm{Q}(\sqrt{D})$. This is a degree 2 number field (since $\sqrt{D}$ is the solution of a degree 2 polynomial $u^{2}+D=0$).

Let us now switch to elliptic curves: A generic elliptic curve $E$ over $\mathbbm{C}$ maps onto $\mathbbm{C}/\Lambda$ where $\Lambda$ is a $\mathbbm{Z}^2$ lattice, $\Lambda=\mathbbm{Z}^2\oplus\tau\mathbbm{Z}$. The endomorphisms of the elliptic curve to itself that fix the origin are holomorphic maps of the form $w\mapsto A w$, where $A\,\Lambda \subset \Lambda $. Any elliptic curve with coordinate $w\in\Lambda$ has an endomorphism that is given by just multiplication with an integer $A\in\mathbbm{Z}$, which corresponds to integer scalings of the lattice. Hence for a generic torus we have  $\text{End}(E)=\mathbbm{Z}$. However, if we pick a lattice generated by $1$ and $\tau$ where $\tau\in\mathbbm{Q}(\sqrt{D})$, we have an additional endomorphism given by multiplication with $A=\omega=\frac{D+\sqrt{D}}{2}$, and $\text{End}(E)=\mathbbm{Z}\oplus a \tau\mathbbm{Z}$ (where $a$ is the coefficient in front of the quadratic term in~\eqref{eq:QuadraticForm}). Such elliptic curves have complex multiplication by $\omega$: Let $A \in \text{End}(E)$, and say the lattice is generated by $\alpha$ and $\beta$ (or equivalently by $1$ and $t=\beta/\alpha$). Then we have
\begin{align}
\label{endo2}
A \alpha=j\alpha+k\beta\,,\qquad A \beta=m\alpha+n\beta\,.
\end{align}
This linear set of equations will have a non-trivial solutions for $(\alpha,\beta)$ if the determinant of the associated matrix is 0, which means that
\begin{align}
\label{lambsol}
A^{2}-(j+n)A+(jn-km)=0\,.
\end{align}

Let us first assume $A \in \mathbbm{R}$. Since $\alpha$ and $\beta$ are linearly independent over $\mathbbm{R}$, the first equation in \eqref{endo2} gives
\begin{align}
\label{endo3}
(A-j)\alpha-k\beta=0\,,
\end{align}
which implies $A \in \mathbbm{Z}$ and leads to the trivial endomorphisms corresponding to lattice rescaling that exist for any elliptic curve. In contrast, if $A$ is allowed to take complex values, we find the solutions
\begin{align}
A=\frac{(j+n) \pm \sqrt{(j-n)^{2}+4km}}{2}
\end{align}
where $D=(j-n)^{2}+4km<0$ (taking $(j-n)^{2}+4km\geq0$ reduces to the case $A \in \mathbbm{R}$). Inserting this into \eqref{endo3} and dividing by $\alpha$, we find (focusing on $t$ in the upper complex half-plane)
\begin{align}
t=\frac{(n-j) + \sqrt{(j-n)^{2}+4km}}{2k}
\end{align}
Hence, $t \in \mathbbm{Q}(D)$ and the endomorphism ring is enlarged. 

The point is now that both the condition~\eqref{lambsol} on $A$ and~\eqref{taucross} for level crossing along the real codimension~1 path in complex structure moduli space parameterized by $\psi\in\mathbbm{R}$ as drawn in Figure~\ref{fig:ModuliSpaceMetric} are a quadratic equation for a single variable ($A$ and $\tau$, respectively) with integer coefficients (the winding numbers around the periods $\alpha$ and $\beta$ or around 1 and $\tau$). The $\text{SL}(2,\mathbbm{Z})$ transformations that leave this quadratic equation invariant are the usual $\text{SL}(2,\mathbbm{Z})$ transformations of the target space torus. Hence, each level-crossing point along the line $\text{Im}(\psi)=0$ is a CM point. Of course, there are many more CM points and many more crossings (for arbitrary complex $\psi$), in particular whenever $\tau_1\in\mathbbm{Q}$.

\subsubsection{Generalization of CM for Calabi--Yau n-folds}
Complex Multiplication for higher-dimensional Calabi-Yau manifolds have been studied in~\cite{borcea1998calabi,Gukov:2002nw}. For higher-dimensional tori, $T^{2n}$, the generalization is rather straight-forward. Let $Z$ be the set of complex coordinates $z_{i}$, $i=1,\cdots , n$. An endomorphism $A$ of $T^{2n}$ is given by
\begin{align}
Z \rightarrow AZ\,.
\end{align}
For $T^{2n}$, we have a set of $n$ 1-forms $\omega_{i}$, and $2n$ 1-cycles $A_{i}$ and $B_{i}$ for $i=1,\cdots,n$. We can set the integrals over the $A$ cycles to 1, and then define the period matrix $T$ as
\begin{align}
T_{ij}=\int_{B_{i}} \omega_{j}\,.
\end{align}
Following the same line of arguments as in the previous section, there will be non-trivial endomorphisms if the equation
\begin{align}
\label{endtor}
TNT+TM-N'T-M'=0
\end{align}
has non-trivial solutions, where $M,N,N',M'$ are integer matrices and $\text{rank}(N)=n$. In that case, the endomorphism $A$ is given by
\begin{align}
\label{endotor1}
A=M+NT\,.
\end{align}

The generalization to Calabi--Yau $n$-folds is to impose the condition~\eqref{endtor} on the middle cohomology, where $T$ are the Calabi-Yau periods computed with respect to the holomorphic $(n,0)$ form $\Omega$ and an integral basis of $n$-cycles.

\subsubsection{The Quartic K3 surface }
\label{sec:CMQuartic}
After the torus, the next-higher one-parameter family of Calabi-Yau manifolds is a K3, given by zero locus of a quartic equation in $\mathbbm{P}^3$. The criteria for a K3 surface $X$ to be CM is described by Chen in~\cite{Chen:2007}. The author studies non-trivial endomorphisms of $H^{2}(X,\mathbbm{Z})$ and show that they occur when the transcendental lattice $T$ of the K3 is of CM-type. This means that the K3 has Picard rank $20$. Such K3 surfaces have been classified by Shioda and Inose~\cite{Shioda:1977aaa} and are called singular (``singular'' means exceptional in that context; the manifolds are smooth) or attractive (since they are related to attractors) K3 surfaces. The construction proceeds by constructing all attractive K3s as a double cover of the orbifold $(T^2\times T^2)/\mathbbm{Z}_2$, and the CM property is inherited from the CM property of the underlying two-tori. One can see that attractive K3 are CM as follows~\cite{Chen:2007}: for Picard rank 20, the transcendental lattice $T$ is
\begin{align}
\label{eq:SplittingK3}
T=H^{2,0}(X)\oplus H^{0,2}(X)\,.
\end{align}
Since a K3 surface has a unique holomorphic 2-form, $T$ is a 2-dimensional vector space. Now we will show that this decomposition is defined over a quadratic CM-field. We can choose an orthogonal basis $(e_{1},e_{2})$ for $T$ over $\mathbbm{Q}$. Next, we can define a basis vector $\sigma$ of $H^{2,0}(X)$ as
\begin{align}
\sigma=e_{1}+A e_{2}\,,
\end{align}
where $A \in \mathbbm{C}$. One can then use the Mukai pairing, which is proportional to the wedge product on differential forms, to show that
\begin{align}
\label{eq:sigma}
0=\langle \sigma ,\sigma \rangle =\langle e_{1} ,e_{1} \rangle+A\langle e_{2} ,e_{2} \rangle\,, \qquad
A^{2} =-\frac{\langle e_{1} ,e_{1} \rangle}{\langle e_{2} ,e_{2} \rangle}<0
\end{align}
Hence, we see that that $\sigma \in \mathbbm{Q}(A)$, and $\mathbbm{Q}(A)$ is a CM-field. Now, there is an endomorphism of $T$ given by the action of $\mathbbm{Q}(A)$,  $A:T\rightarrow T$, which induces the basis change
\begin{align}
e_{1}\rightarrow A^{2}e_{2}\,, \qquad e_{2}\rightarrow e_{1}\,.
\end{align}
Inserting this in~\eqref{eq:sigma}, we get
\begin{align}
A:\sigma\rightarrow A\sigma\,.
\end{align}
This preserves $H^{2,0}(X)$ as was claimed above. Hence, K3 surfaces of Picard rank 20 admit non-trivial endomorphisms of its middle cohomology, corresponding to multiplication by elements in an imaginary quadratic field.

The one-parameter family of quartics in $\mathbbm{P}^3$ has Picard rank 19 generically, but the rank of the Picard lattice is enhanced to 20 over a dense set of countably many points in the moduli space of $\psi$~\cite{Huybrechts:2016uxh}. This begs the question of whether there will be crossings in the spectrum of the Laplacian at these points of enhanced symmetry as well, just like the torus case. The Picard-Fuchs system of the quartic is also hypergeometric, with the fundamental period given by
\begin{align}
\label{3F2}
\varpi_1={}_3F_2\left(\frac{1}{2},\frac{1}{4},\frac{3}{4};1,1;\frac{1}{\psi^{4}}\right)
\end{align}
This K3 is CM at special points, as can be seen by writing the fundamental period in terms of a modular parameter $\tau$~\cite{Dembele:2022aaa},
\begin{align}
\label{modid}
\;{}_3F_2\left(\frac{1}{2},\frac{1}{4},\frac{3}{4};1,1;\rho(\tau)\right)=\bigg(\frac{\eta(\tau)^{16}}{\eta(2\tau)^{8}}+64\frac{\eta(2\tau)^{16}}{\eta(\tau)^{8}}\bigg)^{1/2}\,,
\end{align}
where the function $\rho(\tau)$ is given in terms of Dedekind $\eta$-functions,
\begin{align}
\label{modid2}
\rho(\tau)=\frac{256\,\eta(\tau)^{24}\,\eta(2\tau)^{24}}{(\eta(\tau)^{24}+64\eta(2\tau)^{24})^{2}}\,.
\end{align}
The CM points correspond to choosing a value of $\tau$ where the corresponding torus has complex multiplication. We can then simply calculate the corresponding $\psi$ values from $\psi^{-4}=\rho(\tau)$.

\subsubsection{The Quintic CY threefold}
\label{sec:CMQuintic}
The final case we will discuss is the quintic Calabi-Yau threefold. The period matrix can be obtained from the prepotential $F$ by
\begin{align}
T_{ij}=\partial_{i}\partial_{j}F\,,
\end{align}
where $i,j=0 \cdots h^{2,1}$. Borcea proved that the existence of CM for a CY threefold is equivalent to the condition that the elements of the endomorphism matrix $A$ generate an imaginary field $K$:
\begin{align}
K=\text{End}(H^{3}(M,\mathbbm{Q}))\otimes \mathbbm{Q}\,.
\end{align}
Since $A$ is related to the period matrix $T$ through \eqref{endotor1}, $T$ is also valued in the above number field. Borcea also showed that this number field must have degree $2(h^{2,1}+1)$. Gukov and Vafa~\cite{Gukov:2002nw} checked this for case of the Fermat quintic for $\psi=0$, i.e., they showed that $\psi=0$ is a CM point. In this case, $h^{2,1}=1$, and the extension should be of degree 4. Indeed, the required extension is by the fifth root of unity $\zeta=e^{\frac{2 \pi i}{5}}$, which solves the degree 4 equation
\begin{align}
x^{4}+x^{3}+x^{2}+x+1=0\,.
\end{align}
The period matrix elements $T_{ij}$ takes values in this extension
\begin{align}
\label{quinend}
T=\left(\begin{array}{ccc}
\zeta -1 &  \zeta+\zeta^{3} \\
 \zeta+\zeta^{3} & -\zeta^{4}  \\
\end{array}\right)\,.
\end{align}
One can show that \eqref{quinend} satisfies \eqref{endtor} for specific integer matrices, which lead to the endomorphism
\begin{align}
\label{quinend2}
A=\left(\begin{array}{ccc}
\zeta -1 &  \zeta+\zeta^{3} \\
 1+\zeta+\zeta^{3} & -\zeta^{4}  \\
\end{array}\right)\,.
\end{align}
Hence, the elements of $T$ and $A$ are both valued in $K=\mathbbm{Q}(\zeta)$, in accordance with Borcea. As we will see in our numerical analysis, there are again level crossings at the CM point $\psi=0$. Beyond the Fermat point $\psi=0$, no other CM points are known for the quintic, and there is some numerical evidence that none exist, as we will discuss next in the context of attractors.

\subsection{Attractor points}
\label{sec:Attractors}
The attractor mechanism was introduced in~\cite{Ferrara:1995ih} and states that vector multiplets coupled to spherical, dyonic black holes at the origin of spacetime with a metric
\begin{align}
ds^2=-e^{2U(r)}dt^2+e^{2U(r)}d\vec{x}^2
\end{align}
flow to fixed points in their target space. Here, $U(r)$ is a function of the radial distance $r=|\vec{r}|$. Asymptotically flat space then requires $U(r)\to0$ as $r\to\infty$.

In the context of IIB string compactifications on Calabi-Yau manifolds $X_3$ to 4D (including cases $X_3=X_{3-d}\times T^{2d}$ where the holonomy is a proper subset of SU(3)), the vector multiplet moduli space is associated with the complex structure moduli space on $X_3$ and the charge lattice is given by the middle cohomology lattice, $\Lambda=H^3(X,\mathbbm{Z})$. The fixed points of the attractor flow for a black hole with charges $\hat\gamma\in H^3(X,\mathbbm{Z})$ turns into a condition on the Hodge structure of $X$, 
\begin{align}
\label{eq:HodgeDecomposition}
\hat\gamma=\hat\gamma^{3,0}+\hat\gamma^{0,3}\,.
\end{align}
The (normalized) central charge of the BPS black hole is given by
\begin{align}
\label{eq:BPSCharge}
|Z(\Omega(\psi),\gamma)|^2=\frac{\left|\int_\gamma \Omega(\psi)\right|}{i\int\Omega(\psi)\wedge\bar\Omega(\psi)}\,,
\end{align}
where $\Omega(\psi)$ is the complex-structure moduli-dependent holomorphic top-form and $\gamma$ is Poincar\'e dual to $\hat\gamma$. In~\cite{Ferrara:1997tw}, it was shown that stationary points of $|Z|^2$ with $Z\neq0$ occur if and only if $\hat\gamma$ has the decomposition~\eqref{eq:HodgeDecomposition}.

In his study of attractors, Moore distinguishes three cases:
\begin{enumerate}
\item $|Z(\psi_{*},\gamma)|\neq0$: The flow exists for any distance $r$ from the BPS object.
\item $|Z(\psi_{*},\gamma)|=0$ and $\psi_{*}$ is at a regular point in moduli space: The flow breaks down at finite distance and no BPS object exists.
\item $|Z(\psi_{*},\gamma)|=0$ and $\psi_{*}$ is at a singular or boundary point in moduli space: Well-behaved solutions might or might not exist. An example of a well-behaved point with this property is the conifold point of the quintic.
\end{enumerate}

As explained in~\cite{Denef:2001xn}, the attractor flow equations for $U$ and\footnote{Remember that a good coordinate on complex structure moduli space of the family of CY manifolds in~\eqref{eq:DworkCY}is $\psi^{n+1}$.} $z=\psi^{-(n+1)}$ in terms of the inverse radial distance $\tau=r^{-1}$ reads
\begin{align}
\label{eq:AttractorFlow}
\partial_\tau z = -g^{z\bar z} \bar\partial_{\bar z}|Z|\,,
\end{align}
where $g^{z\bar z}>0$ is the inverse moduli space metric. By redefining $\tau$, we can absorb $g^{z\bar z}$ in $\tau$, which further simplifies the flow equation. This makes it obvious that the flow ends at stationary points of $|Z|$. 

To compute the attractor points as an end point of the flow, one proceeds as follows: In the mirror-dual IIA theory, the (quantum-corrected) volume of holomorphic even-dimensional cycles are defined in terms of the BPS D-brane states $(D_0,D_2,D_4,D_6)$ that wrap these cycles. These quantum-corrected volumes on the IIA side are given on the IIB side in terms of the mirror 3-cycle. Choosing an integral  basis of three-cycles $\Gamma_i$, $i=1,\ldots 2(h^{2,1}+1)$ of $H^3$ such that $\gamma=q^i\Gamma_i$, one can express~\eqref{eq:BPSCharge} as
\begin{align}
|Z(\gamma)|=\frac{\left|q^i\int_{\Gamma_i}\right|}{\left[i\int_X\Omega\wedge\bar\Omega\right]^{1/2}}\,.
\end{align}
The periods of the one-parameter family of quintics has been computed already in the early days of mirror symmetry~\cite{Candelas:1990rm}. In terms of the period vector $\Pi$, we can write the (normalized) central charge simply as
\begin{align}
\label{eq:CentralCharge}
Z(q;z,\bar z)=e^{\mathcal{K}(z,\bar z)/2}\;q\cdot\Pi\,, \qquad \mathcal{K}(z,\bar z) = -\ln\left[ i\int_X \Omega\wedge\bar\Omega\right]\,.
\end{align}

From the above discussion, the condition on attractors and CM look similar, compare e.g.~\eqref{eq:HodgeDecomposition} and~\eqref{eq:SplittingK3}. In~\cite{Moore:1998pn} Moore observed and studied the connection between complex multiplication and attractor points. Indeed, via the Shioda-Inose construction~\cite{Shioda:1977aaa}, there is a one-to-one correspondence between attractive K3 surfaces and CM tori. Moreover, Moore conjectured that all attractors are arithmetic, i.e., the periods and the complex structure coordinates are both arithmetic. Moreover, he observed that compactification of F-theory on attractive K3 surfaces corresponds under the usual heterotic F-theory duality to the 8D heterotic string being a Rational Conformal Field Theory (RCFT). Gukov and Vafa conjectured~\cite{Gukov:2002nw} that SCFTs whose target space are K3 surfaces with CM are themselves rational SCFTs.

While the connection between CM points, attractors and arithmeticity is well established for CY one- and two-folds, it is false in general. In~\cite{Lam:2020qge}, the authors construct counterexamples for CY $n$-folds with odd $n$ (except for $n=1,3,5,9$). But already on CY threefolds, it is not expected that attractors are arithmetic in general. Gukov and Vafa already point out that CM and hence RCFTs are expected to appear at most at a finite number of points for the 1 parameter family of quintics, based on the Andr\'e-Oort conjecture~\cite{Andre:1989aaa,Oort:1997aaa} (which has been recently been proven~\cite{Pila:2022aaa}). Moreover, the authors of~\cite{Candelas:2019llw} study this family of quintics and search for rank 2 attractor points, which are attractor points for two independent integral charges $\gamma_1$ and $\gamma_2$. These are arithmetic by the Hodge conjecture, but the rank one attractor condition seems not strong enough to guarantee arithmeticity. Numerical evidence presented in~\cite{Candelas:2019llw} indeed suggests that the periods are not algebraic but transcendental.

Given that points in complex structure moduli space where eigenvalues cross are ubiquitous and CM points are rare beyond K3, this makes a correspondence between crossings and CM unlikely. However, attractor points are also ubiquitous in these compactifications~\cite{Denef:2001xn}. Hence, a more promising generalization from the torus and K3 seems to be that crossings are not related to CM but to (rank 1) attractor points. For the one parameter family of $T^2$ and $K3$ manifolds the CM and attractor conditions are the same, so we can investigate crossings and CM points for the torus and K3, and move to analyzing attractor points for the quintic threefold, which we do in the following section. Our numerical studies supports a relation between attractor points and level crossing. It would be very interesting to understand this connection on a deeper level, since attractor points are related to BPS objects, while we are not aware for such a relation for higher eigenmodes of the Laplacian.

\section{Numerical spectrum}
\label{sec:NumericSpectrum}
In order to check whether level crossing of the different modes of the scalar Laplace operator correspond to attractor (or CM) points for CY $n$-folds, we need to resort to numerical methods. The purpose of this section is to study potential sources of inaccuracies in the numerical spectrum computation of the Laplace operator.

The spectrum computation proceeds in several steps, each of which requires some approximation and hence introduces numerical inaccuracies:
\begin{enumerate}
\item First, instead of working over the entire complex variety $X$, we sample a collection of points and evaluate the quantities of interest on these points. The denser the point sample, the more accurate this approximation becomes.
\item Second, the Ricci-flat Calabi-Yau metric has to satisfy a complex differential equation of Monge-Ampere type~\cite{Calabi:1957aaa, Yau:1977ms}. We approximate the exact metric using a neural network (NN), which we evaluate at the points sampled in step 1. The accuracy will improve if we include more points when training the neural network, when we make the NN bigger (i.e., a more powerful function approximator), and when we train the NN longer (i.e., improve the accuracy with which the NN satisfies the Monge-Ampere equation).
\item Third, we construct a set of basis functions in which we expand the eigenfunctions of the scalar Laplace operator. We construct this basis from pullbacks of the basis functions that parameterize the lowest eigenmodes of the scalar Laplace operator in the ambient space $\mathbbm{P}^{n+1}$. By including more basis functions, we can improve the approximation capacity of the eigenmodes on the CY. In particular, we would expect that the accuracy of the higher modes on the CY suffers from this truncation.
\end{enumerate}

In order to benchmark our pipeline, i.e., to gauge the influence of the inaccuracies introduced in each of the steps on the final spectrum, we compare the numerical and the analytic, exact results for $T^2$ as we vary the parameters controlling the several truncations described above. Once we have a qualitative idea of which factors influence the result the strongest and how big the inaccuracies are, we check whether level crossings occur at the CM points discussed in Section~\ref{sec:CMQuartic} for the quartic K3 and Section~\ref{sec:CMQuintic} for the quintic CY threefold. Knowledge of the influence of these hyperparameters on the spectrum approximation will also be useful for other studies and applications that involve the higher Laplacian eigenmodes.

\subsection{Point generation}
To generate points on the Calabi-Yau which are distributed according to the pullback of the Fubini-Study (FS) metric on the ambient $\mathbbm{P}^n$, we use the techniques detailed in~\cite{Anderson:2020hux} and implemented in the cymetric package~\cite{Larfors:2021pbb, Larfors:2022nep}. The points sampled will lie on the CY $X$, but are expressed in terms of the homogeneous coordinates on the ambient $\mathbbm{P}^2$. To plot the distribution, we map the Fermat cubic~\eqref{eq:cubic} into Weierstrass form~\eqref{eq:weierstrassform} using~\eqref{weimap}, use the projective scaling to set $z=1$ (for points sampled numerically, this is always possible, since the points that have $z=0$ are a measure zero set and will hence never be sampled), invert the Weierstrass $\wp$ function to obtain the corresponding value in terms of the ``flat'' coordinate $w\in\mathbbm{C}/\Lambda$, and use lattice translations to map the points into the fundamental cell spanned by $1$ and $\tau(\psi)$, as calculated from inverting~\eqref{eq:Kleinj}.

\begin{figure}[t]
\centering
\begin{tabular}{lll}{\includegraphics[width=0.27\textwidth]{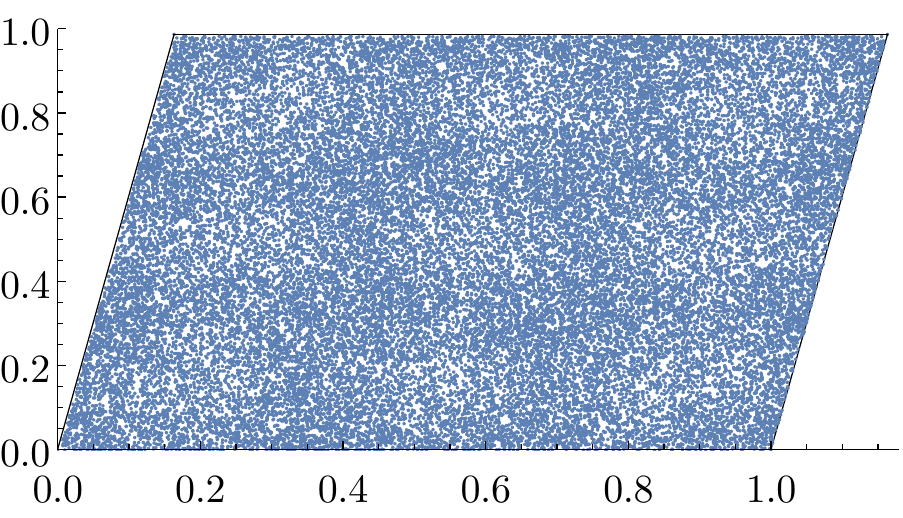}}\quad&
{\includegraphics[width=0.27\textwidth]{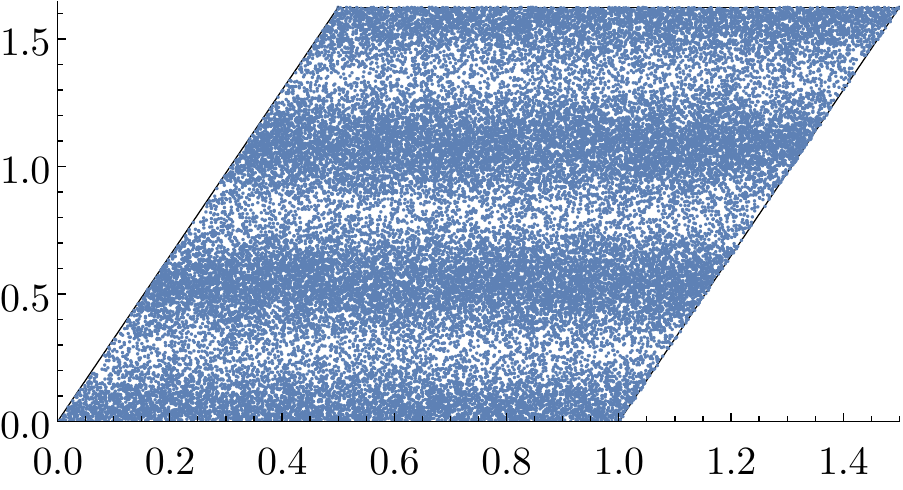}} \quad&
{\includegraphics[width=0.27\textwidth]{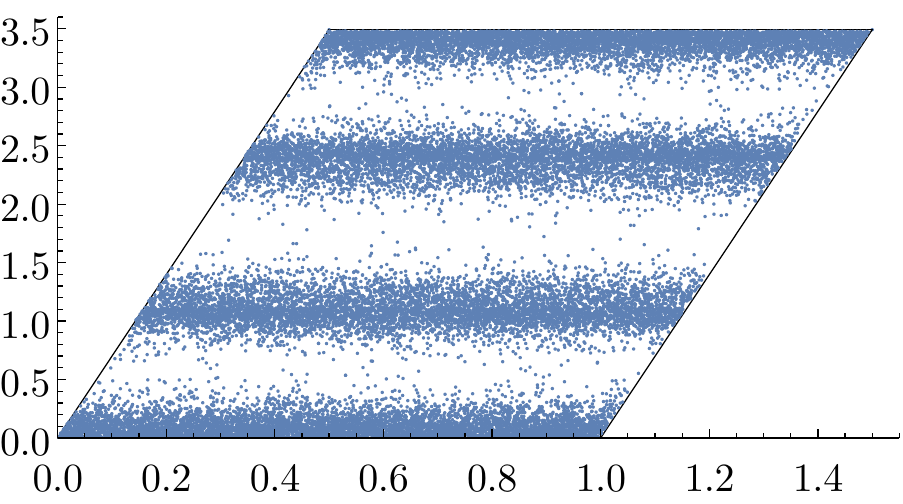}} \\
\subfigure[$\psi=-1$, $\tau=0.16 + 0.99 i$]
{\includegraphics[width=0.3\textwidth]{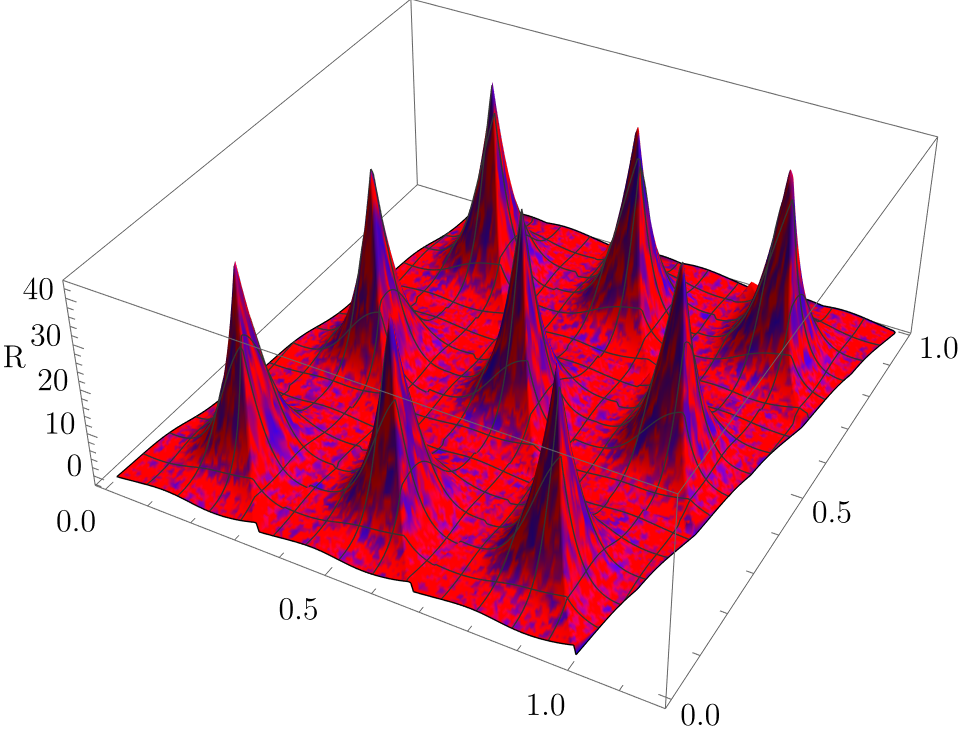}}\quad&
\subfigure[$\psi=-10$, $\tau=0.5 + 1.62 i$]
{\includegraphics[width=0.3\textwidth]{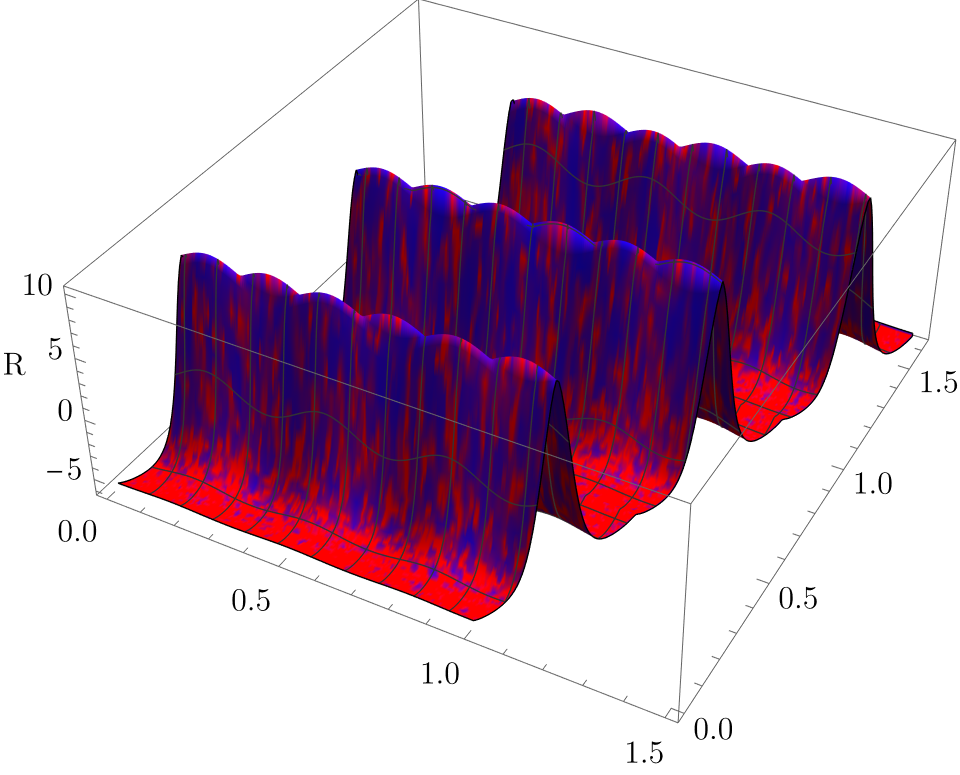}} \quad&
\subfigure[$\psi=-500$, $\tau=0.5 + 3.5 i$]
{\includegraphics[width=0.3\textwidth]{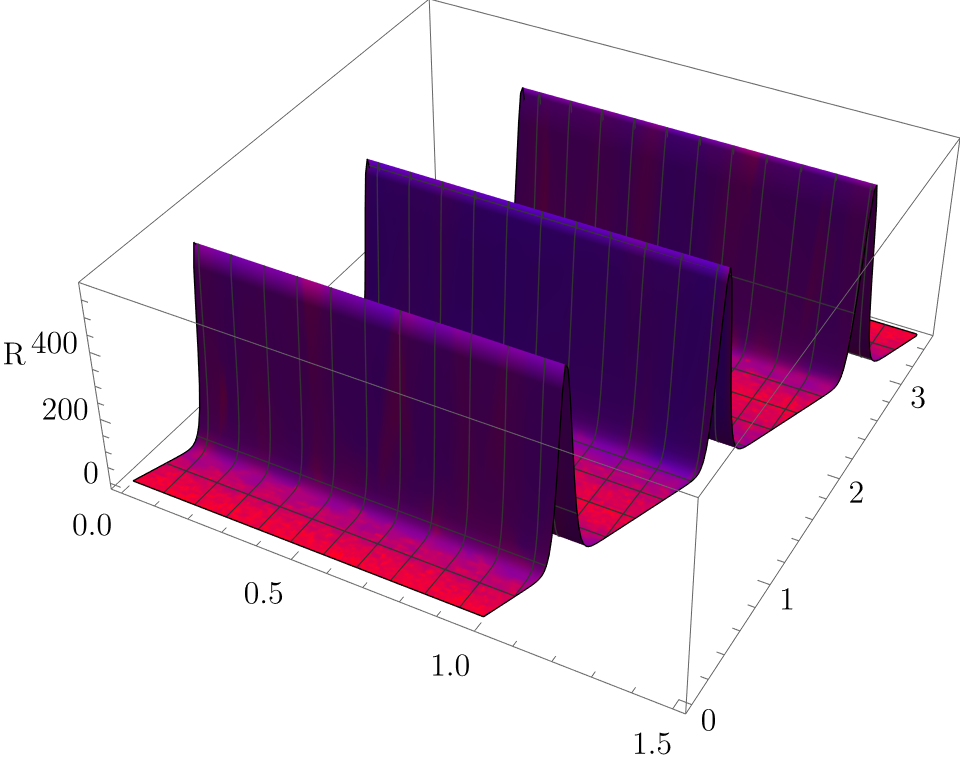}}
\end{tabular}
\caption{Top row: Distribution of points sampled from the ambient space with respect to the pullback of the FS metric for various $\psi$. We see horizontal voids develop for larger $|\psi|$. Bottom row: The Ricci scalar $R$ (w.r.t.~ the FS metric) in the fundamental domain, with color gradient indicating the point density. Higher curvature corresponds to undersampled regions.} 
\label{fig:pointdist}
\end{figure}

In Figure~\ref{fig:pointdist}, we plot the distribution of 40000 point samples for $\psi\in\{-1,-10,-500\}$. Note that, as $|\psi|$ gets larger, horizontal voids develop. These hint at the point sampling method sampling some regions on the Calabi-Yau more densely than others. The sampling technique we use produces points that are distributed according to the pullback of the Fubini-Study metric, so these voids are an artifact of how the hypersurface is embedded in the projective ambient space. Since we know the point distribution explicitly, we can correct for over-/undersampling regions by weighting points in oversampled regions less and points in undersampled regions more, according to the ratio of the pullback of the ambient space FS metric and the volume measure on the CY as computed from $|\Omega|^2$, with the weights $w\sim\det(g)/|\Omega|^2$. Indeed, $|\Omega|^2$ is small in the undersampled regions, while the difference in the (determinant of the) of the pullback of the FS metric is not very pronounced. Furthermore, this correlates with the curvature in the void regions (of the pulled back FS metric) being larger than elsewhere, as seen in Figure~\ref{fig:pointdist}.

We can also exclude that the problem is numerical and arises from the hierarchy in the coefficients of the defining polynomials. We do this by using $SL(2,\mathbbm{Z})$ transformations of $\tau$ to identify an equivalent point $\psi'(\tau)$, whose absolute value is close to 1: As can be seen in Figure~\ref{fig:ModuliSpaceMetric}, the infinite complex structure point is mapped to $|\psi|=1$. The fact that the point sample for these numerically very different values of $\psi$ looks identical excludes that the voids are due to numerical problems.

\subsection{Spectrum}
Next, we compute the spectrum numerically, varying the number of points $n_p$, the complex structure parameter $\psi$ and the number $k_{\phi}$ of the basis functions in which we expand the Laplacian eigenfunctions. We perform all computations for the pullback of the FS metric (which is the lowest-order approximation to the CY metric in the sense of Donaldson's algorithm~\cite{Donaldson:2005aaa}) and for the exact CY metric (obtained from $|\Omega|^2$, which is proportional to the determinant of the metric, and hence to the metric itself for one-folds) to see the influence of choosing various qualities of approximations to the CY metric. In all cases, we can compare the approximate result to the analytic result~\eqref{eq:eigenvalues} to quantify the error of the approximation.

\subsubsection*{Varying the number of points}
\begin{figure}[t]
\centering
\includegraphics[width=.95\textwidth]{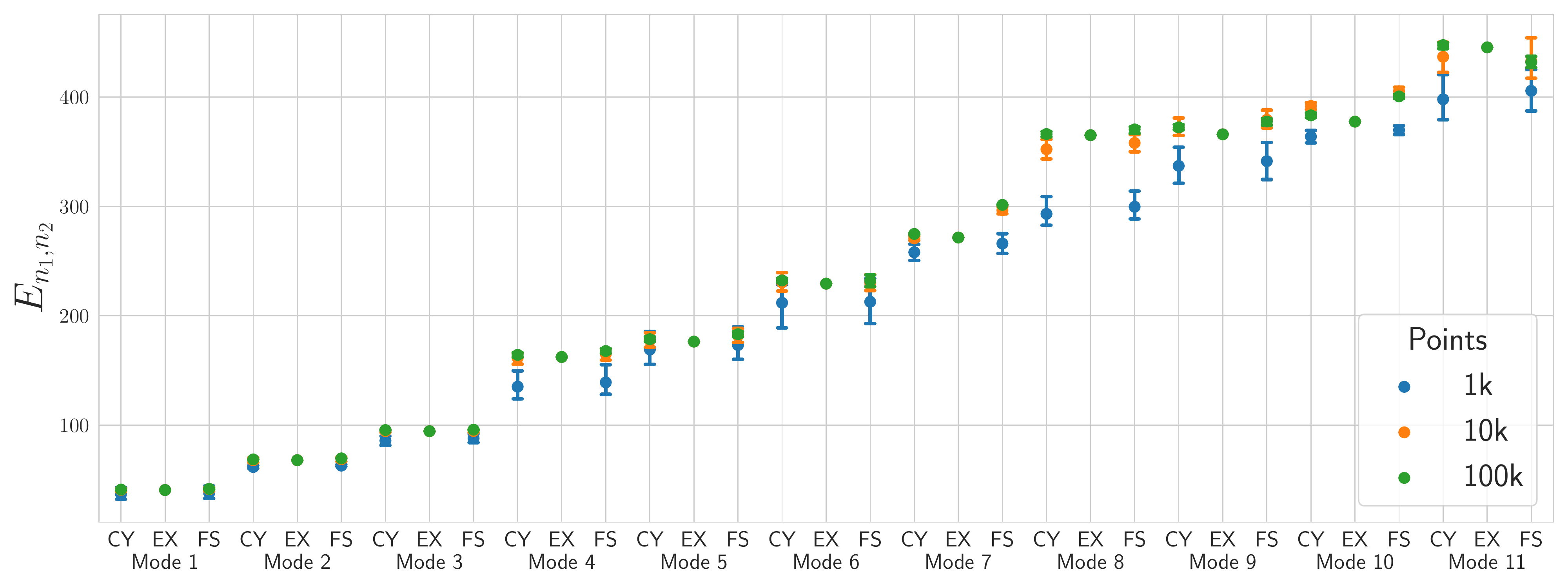}\\[10pt]
\caption{The first 36 massive eigenmodes (averaged per multiplet with error bars corresponding to one standard deviation) as we vary the number of points for the FS and the exact CY metric, compared to the analytic result.}
\label{fig:ErrorsPoints}
\end{figure}

To study the influence of the number of points, we choose $n_p\in\{1,000, 10,000, 100,000\}$. We present the results for each of the first 36 massive eigenmodes (the single massless mode is omitted from the plot) in Figure~\ref{fig:ErrorsPoints}. These 36 eigenmodes fall into various irreps under the symmetry group, such that there are 11 distinct eigenvalues. For each eigenvalue, we plot the spectrum as computed with respect to the exact CY metric obtained from $|\Omega|^2$ (labeled CY in the plot), the analytic result computed from~\eqref{eq:eigenvalues}, and the spectrum computed on the CY hypersurface when using the pullback of the ambient space FS metric as a proxy for the exact CY metric. For the plot, we fix the other parameters like $k_\phi=3$ and $\psi=-1$. The error bars represent 95 percent confidence intervals for multiplets with multiplicity larger 1. The different colors represent the three different choices for the number of points used to compute the spectrum.

From the plot, we can make the following two observations. First, the metric dependence is rather weak. In particular, the error we get from using the FS metric is often comparable to the error we get for the exact CY metric. At around the eighth-heaviest eigenmode, the exact CY metric still agrees very well with the analytic result, while the FS metric result starts to deviate from the other two.

Second, for the lower eigenmodes, as little as 1000 points and the FS metric already gives very good agreement with the exact, analytic result. From the fourth-heaviest multipliet onward, however, the small number of points introduces a significant error, and 10k points are needed for agreement with the exact result. After reaching the eighth heaviest multiplet, 100k points are necessary to obtain agreement with the analytic result.

\subsubsection*{Varying the complex structure}
\begin{figure}[t]
\centering
\includegraphics[width=.95\textwidth]{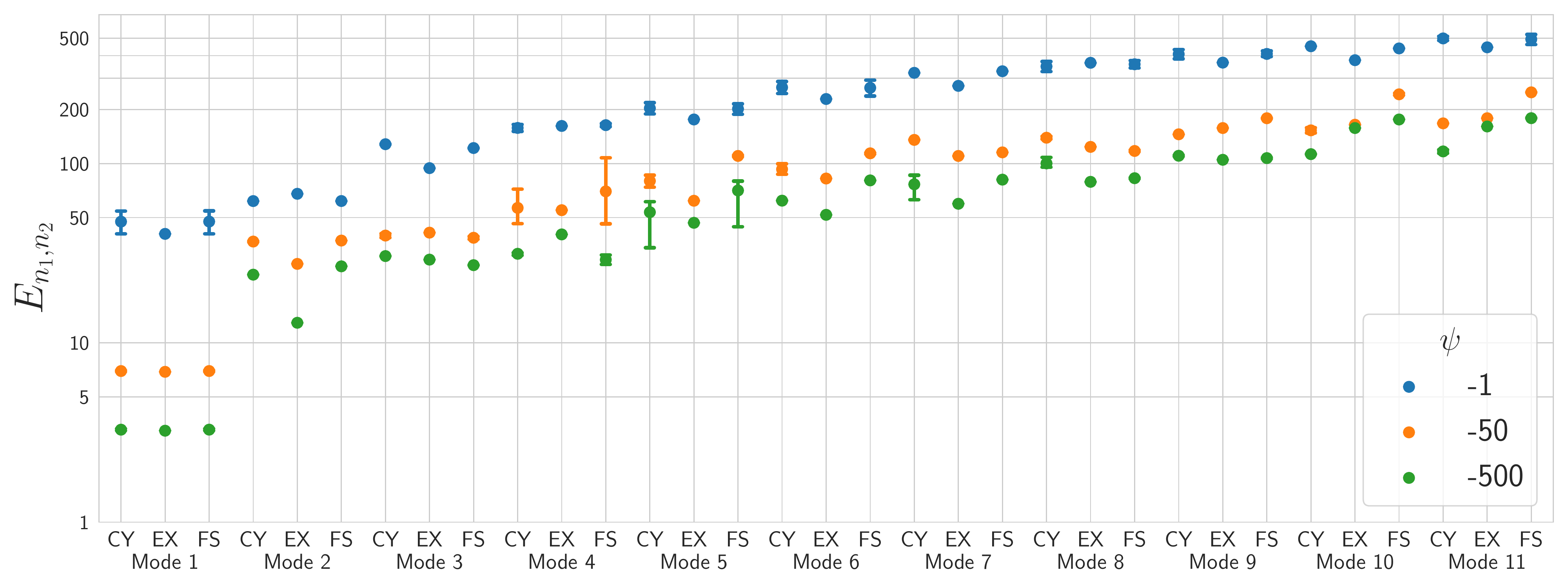}\\[10pt]
\caption{The first 36 massive eigenmodes (averaged per multiplet with error bars corresponding to one standard deviation) as we vary the complex structure for the FS and the exact CY metric, compared to the analytic result.}
\label{fig:ErrorsPsis}
\end{figure}
For the complex structure, we choose $\psi\in\{-1,-50,-500\}$. We focused on this range since we restricted our discussion of CM points to the case where $\text{Im}(\psi)=0$. In terms of the modular parameter $\tau$, these $\psi$ probe a point on the boundary torus (for $\psi=-1$) as well as points along the boundary line with $\tau_1=1/2$, cf.\ Figure~\ref{fig:ModuliSpaceMetric}. We compute the spectrum using 100k points and $k_\phi=3$ for the first 12 multiplets using the exact CY metric as well as the pullback of the FS metric and plot them together with the analytic result in Figure~\ref{fig:ErrorsPsis}. Note that the eigenvalues are complex-structure dependent, so there are three values for each multiplet, one for each choice of $\psi$, which we distinguish by color.

The blue data points with $\psi=-1$ correspond to the green ones in Figure~\ref{fig:ErrorsPoints} and are included for reference (note the logarithmic scale in Figure~\ref{fig:ErrorsPsis} to better resolve the differences for $\psi=-50$ and $\psi=-500$). The trend that lower eigenmodes are approximated better continues to hold also for other values of $\psi$. The dependence on the choice of the metric seems again rather weak. Moreover, the approximation gets worse with larger $\psi$, even for the exact CY metric. The only explanation that we see for this is that even at 100k points, there is too little information in some regions of the CY (the voids) for a reliable computation: The few points that are sampled in the voids are weighted more strongly (by a factor of 10-1000), but there are simply not enough for an accurate spectrum estimation in these regions. We expect that the problem can be overcome by sampling $N_1\gg N$ points initially and then draw a more representative sample by choosing $N$ out of the $N_1$ points according to their weights. Since point generation is rather inexpensive computationally (at least from complete intersections at small $h^{1,1}$), this is a viable option to increase the accuracy of the approximation.

\subsubsection*{Varying the number of basis functions}
\begin{figure}[t]
\centering
\includegraphics[width=.95\textwidth]{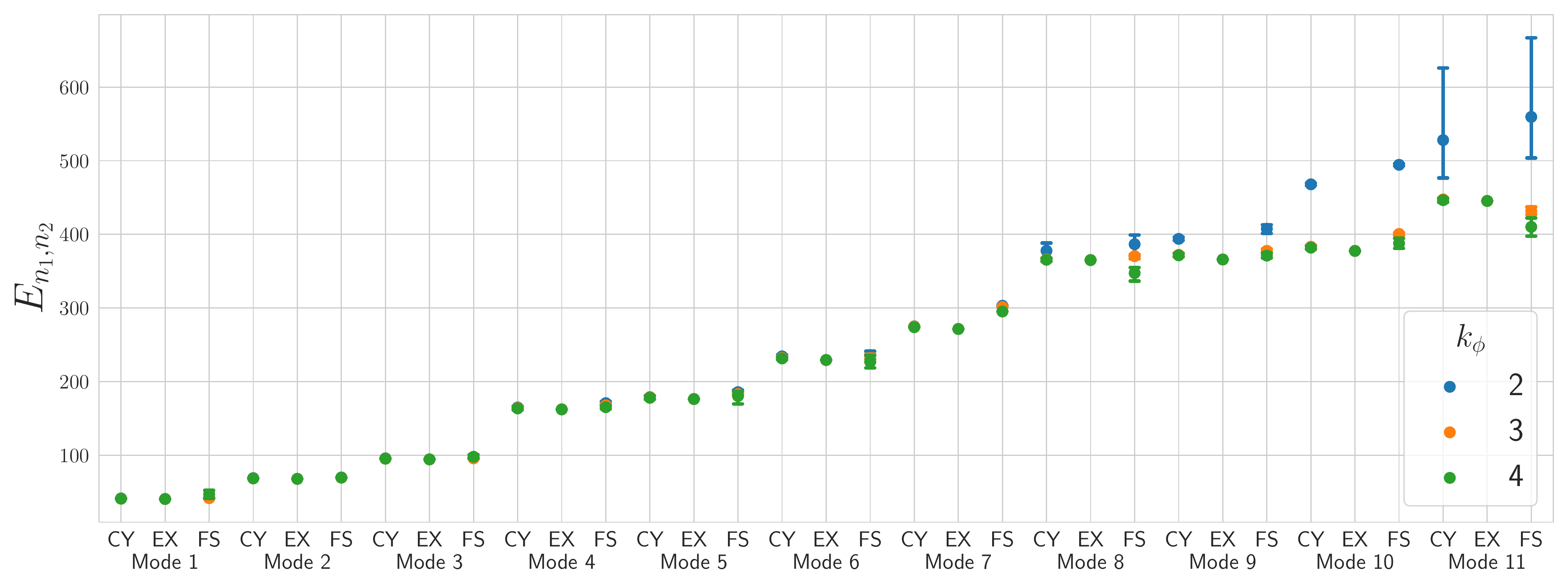}\\[10pt]
\caption{The first 36 massive eigenmodes (averaged per multiplet with error bars corresponding to one standard deviation) as we vary the number of basis functions for the FS and the exact CY metric, compared to the analytic result.}
\label{fig:Errorsks}
\end{figure}
As a set of basis function in which we expand the eigenmodes of the scalar Laplacian on the CY, we choose a set of basis functions (with respect to the FS metric) for the lowest eigenmodes of the scalar Laplacian on $\mathbbm{P}^2$, and pull them back to the Calabi-Yau. In practice, it is to be expected that any of the CY eigenfunctions will receive contributions from all modes of the ambient space Laplacian, so the truncation to a finite set will introduce an error. 

The basis for the ambient space eigenfunctions is
\begin{align}
\{\alpha_{A}\}=\frac{\{s_{\alpha}^{(k_{\phi})} \bar{s}_{\beta}^{(k_{\phi})}\}}{(|z_{0}|^2+|z_{1}|^2+|z_{2}|^2)^{k_{\phi}}}\,.
\end{align}
In this expression, $s_\alpha^{(k_{\phi})}\in H^{0}(\mathcal{O}_{\mathbbm{P}^2}(k_{\phi}))$ are the sections of $\mathcal{O}_{\mathbbm{P}^2}(k_{\phi})$, i.e., monomials of degree $k_{\phi}$ in the homogeneous ambient space coordinates, labeled by $\alpha=1,\ldots,h^{0}(\mathcal{O}_{\mathbbm{P}^2}(k_{\phi}))$. Since there are $n+k_{\phi}$ choose $k_{\phi}$ sections of $\mathcal{O}_{\mathbbm{P}^n}(k_{\phi})$, we find that $A\in\{9,36,100,225\}$ for $k_\phi\in\{1,2,3,4\}$. However, once $k_\phi\geq n+1$ (i.e., the degree of the sections of the anticanonical bundle), the hypersurface equation implies relations among these sections. So, upon pulling back to $X$, only 81 of the 100 ambient space monomials $\alpha_{A}$ at $k_\phi=3$ are independent (and similar for $k_\phi=4$, $144$ of the $225$ are independent). This means that we have in principle access to the lowest 9, 36, 81, and 144 eigenmodes. We would expect that the higher eigenmodes are effected more strongly by the truncation of the number of basis functions. We can see this in Figure~\ref{fig:Errorsks}.

For our error analysis we fix the number of points to 100k and $\psi=-1$. As expected, we find that around the eighth multiplet (which means after 24 eigenmodes), $k_\phi=2$ ceases to provide good approximations, since we are reaching a point where we are using $24/36$ available eigenmodes and the finite truncation effects start to show. At the 36th eigenmode, the full basis for $k_\phi$ has been exhausted and the approximation breaks down. In contrast, both $k_\phi=3$ and $k_\phi=4$ (which comprise the first 100 and 225 ambient space eigenmodes, respectively) are ample to provide an excellent approximation to the spectrum. The metric dependence is again weak, but it is perhaps the most pronounced here: With 100k points at $\psi=-1$ and $k_\phi >2$, the error from the discrete point sampling and from truncating the eigenbasis become negligible, so that the error due to the approximated metric are the main source of discrepancy between the numerical and the exact result. This observation is also supported by the fact that the numerical eigenvalue computation using the exact CY metric is in essentially perfect agreement with the analytic result.

\section{Numerical analysis of crossings and attractor points}
\label{sec:NumericAnalysis}
In this section, we use the numeric methods outlined in Section~\ref{sec:NumericSpectrum} to find crossings of eigenmodes of the scalar Laplacian and compare them to the known CM points on the underlying manifolds.

\subsection{Crossing and CM points for the torus}
For the torus, the analytic results of Section~\ref{sec:XingAndCM} establish that $\psi\in\mathbbm{R}$ gives rise to CM points when two eigenmodes cross. From the discussion of Section~\ref{sec:NumericSpectrum}, we know that we can approximate the eigenspectrum numerically with large accuracy (especially at smaller $\psi$), which allows us the to match the crossings and the CM points at the same level of accuracy.

\subsection{Crossing and CM points for the quartic}
For the quartic, we will exemplify how one can use crossings of eigenmodes in the spectrum to find CM points for
\begin{align}
\tau=\frac{1+3i}{2}\qquad\text{and}\qquad\tau=\frac{1+\sqrt{3}i}{2}\,.
\end{align}
These correspond to 
\begin{align}
\label{eq:CMPointsQuartic}
 \psi^{-4}=\rho(\tau)=-\frac{1}{48}\qquad\text{and}\qquad \psi^{-4}=\rho(\tau)=-\frac{9}{16}\,,
\end{align}
as explained in Section~\eqref{sec:CMQuartic}.

For generic $\psi$, the one-parameter Fermat quartic has a symmetry group 
\begin{align}
(S_{4}) \rtimes (\mathbbm{Z}_{4}\times\mathbbm{Z}_4)\,.
\end{align}
Using the same method that we outlined in Section~\ref{sec:DiscreteSymmetries} for the torus, we find the irreps
\begin{align}
\label{eq:MultiplicitiesK3}
\begin{array}{ |c |c c c c c| } 
 \hline
 \text{dim(irrep)} & 1 & 2 & 3 & 6 & 12\\ 
\hline
\text{number(irreps)} & 2 & 1 & 6 & 1 & 2 \\ 
 \hline
\end{array}
\end{align}

\begin{figure}[t]
\centering
\includegraphics[width=0.8\textwidth]{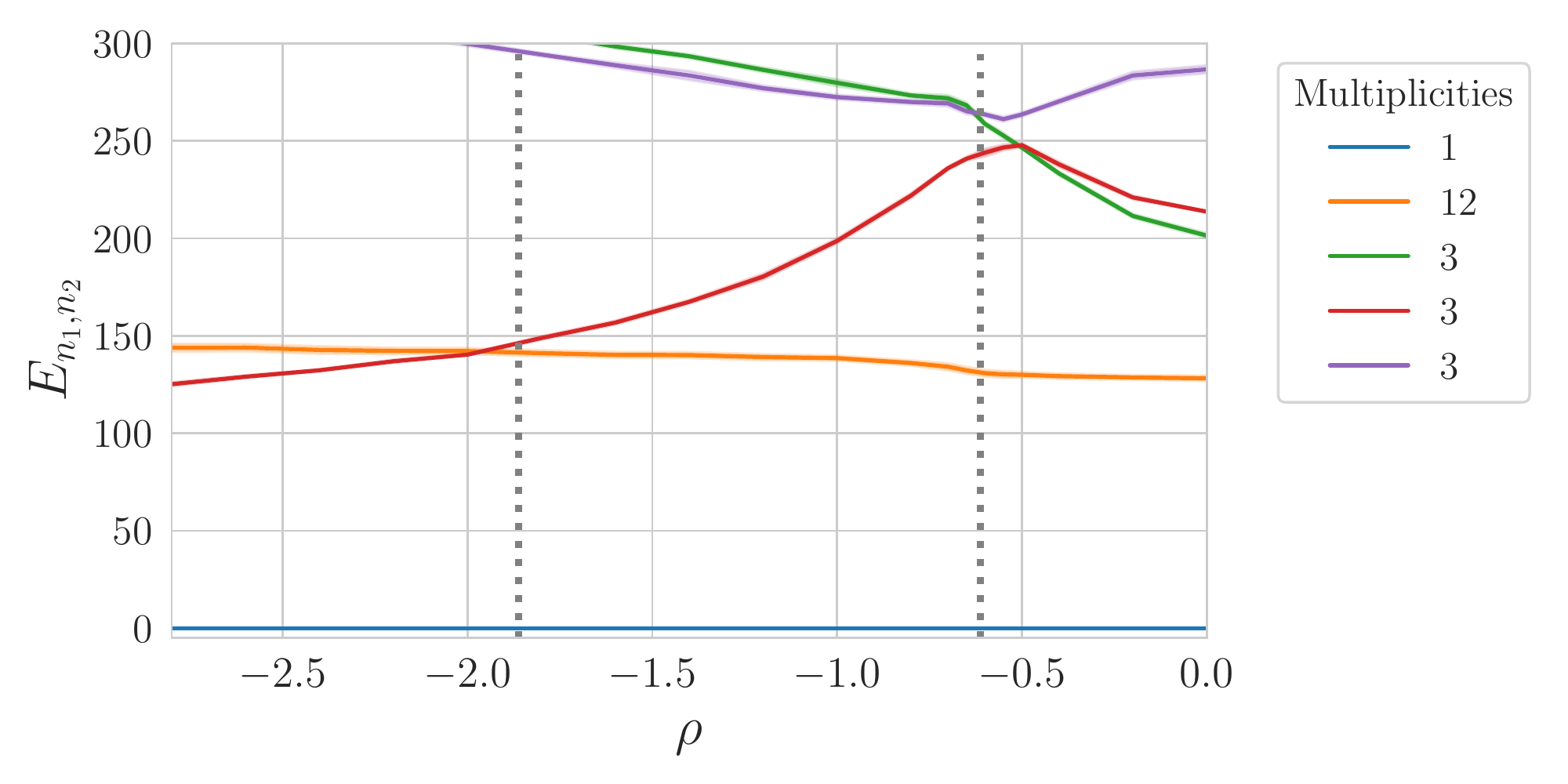}
\caption{Spectrum of the scalar Laplacian on the quartic as a function of complex structure. We plot a codimension 1 slice $\psi=(1+i)\rho$ in the moduli space, which contains CM points at $\psi^{-4}=-1/48$ and $\psi^{-4}=-9/16$ (corresponding to the dashed lines at $\rho=-0.61$ and $\rho=-1.86$). We can see eigenvalue crossings in the vicinity of these values.}
\label{fig:CMQuartic}
\end{figure} 

In our numerical analysis, we choose an arbitrary branch of the fourth root of unity and approach the CM points along the trajectory $\psi=(1+i)\rho$ for $\rho\in\mathbbm{R}$. We approximate the CY metric using the cymetric package~\cite{Larfors:2021pbb,Larfors:2022nep}. We use the phi model with a three-layer neural network (NN) with 64 hidden nodes each and gelu activation, and train the NN with 1 million points generated for $\rho\in[-3,0]$. We train the NN until the sigma loss is $\lesssim 0.01$, which happens already at around 5 epochs. With this approximate CY metric, we then compute the spectrum using $k_\phi=2$, which gives us access to the first 100 eigenmodes of the scalar Laplacian on $\mathbbm{P}^3$. The full calculation takes around 2 hours on a modern desktop PC. We then group the eigenmodes according to their multiplicities as computed in~\eqref{eq:MultiplicitiesK3}. The result is shown in Figure~\ref{fig:CMQuartic}. We see that there are eigenmode crossings among low eigenmodes that are consistent with the CM points on the quartic given in~\eqref{eq:CMPointsQuartic}. We want to point out, however, that the crossing around $\rho=0.61$ is hard to disentangle and could also be consistent with the red, green, and purple line approaching each other but not actually crossing. Since all three have multiplicity 3, they cannot be distinguished by their multiplicity, unlike the much cleaner crossing around $\rho=-1.86$. In any case, the spectrum behaves in a special way around $\rho-=0.61$ as compared to other values for $\psi$, where the eigenmodes just decay or grow exponentially.

\subsection{Crossing and attractor points for the quintic}
For the quintic, the only CM point that is known analytically is at $\psi=0$, Computation of the scalar Laplacian eigenmodes show a plethora of crossings at $\psi=0$, which is known to be a CM point as explained in Section~\ref{sec:CMQuintic}. However, the point $\psi=0$ is special in many regards. In particular, the symmetry group from the ambient space enhances, so this might not be the best point to test the relation. Moreover, as argued in Section~\eqref{sec:Attractors}, it is more likely that the generalization from the torus case is to a connection between eigenvalue crossings and attractor points rather than eigenvalue crossings and CM points.

\begin{figure}
\centering
\begin{tabular}{lll}
\includegraphics[width=.31\textwidth]{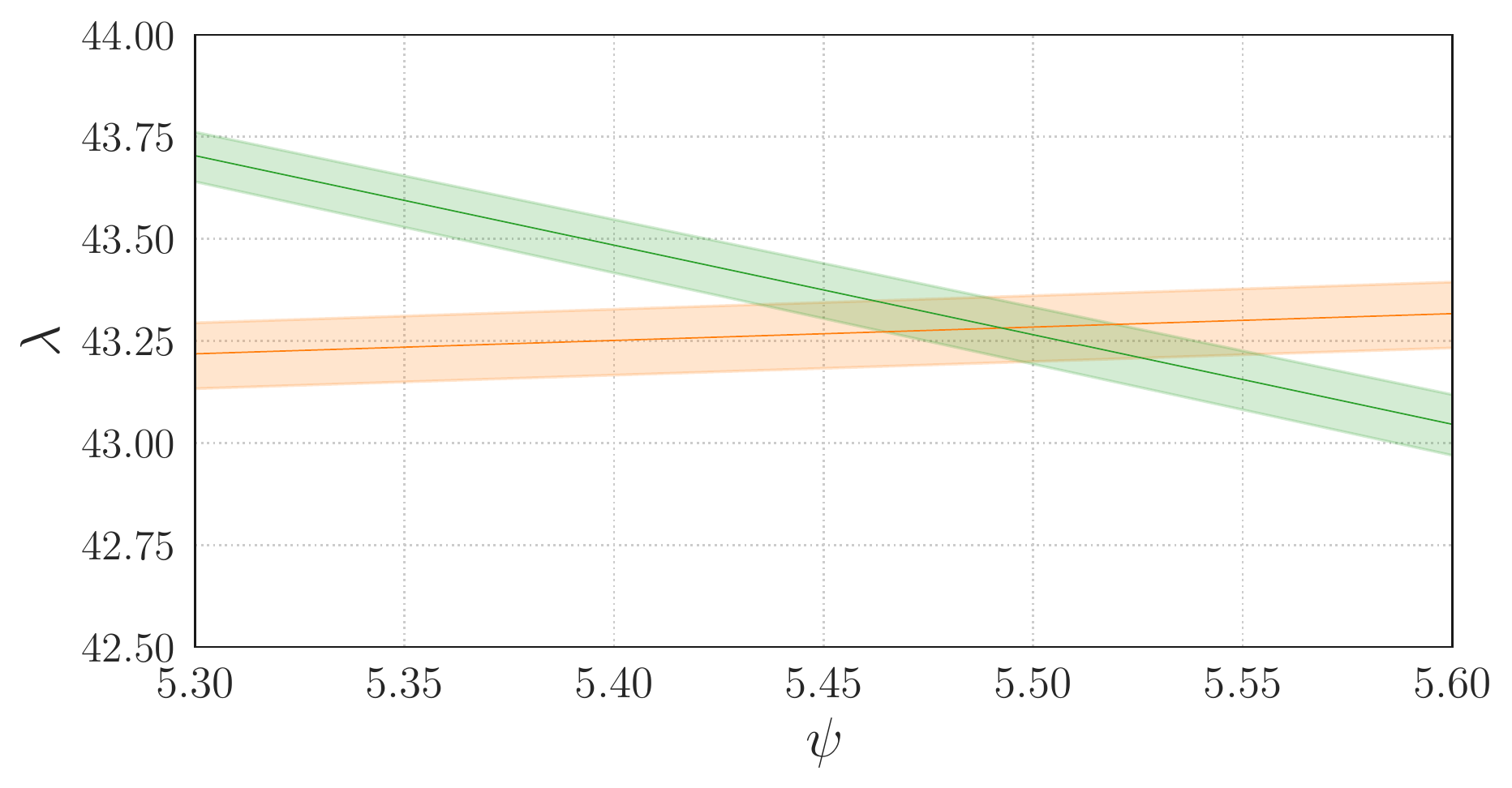}\quad&
\includegraphics[width=.31\textwidth]{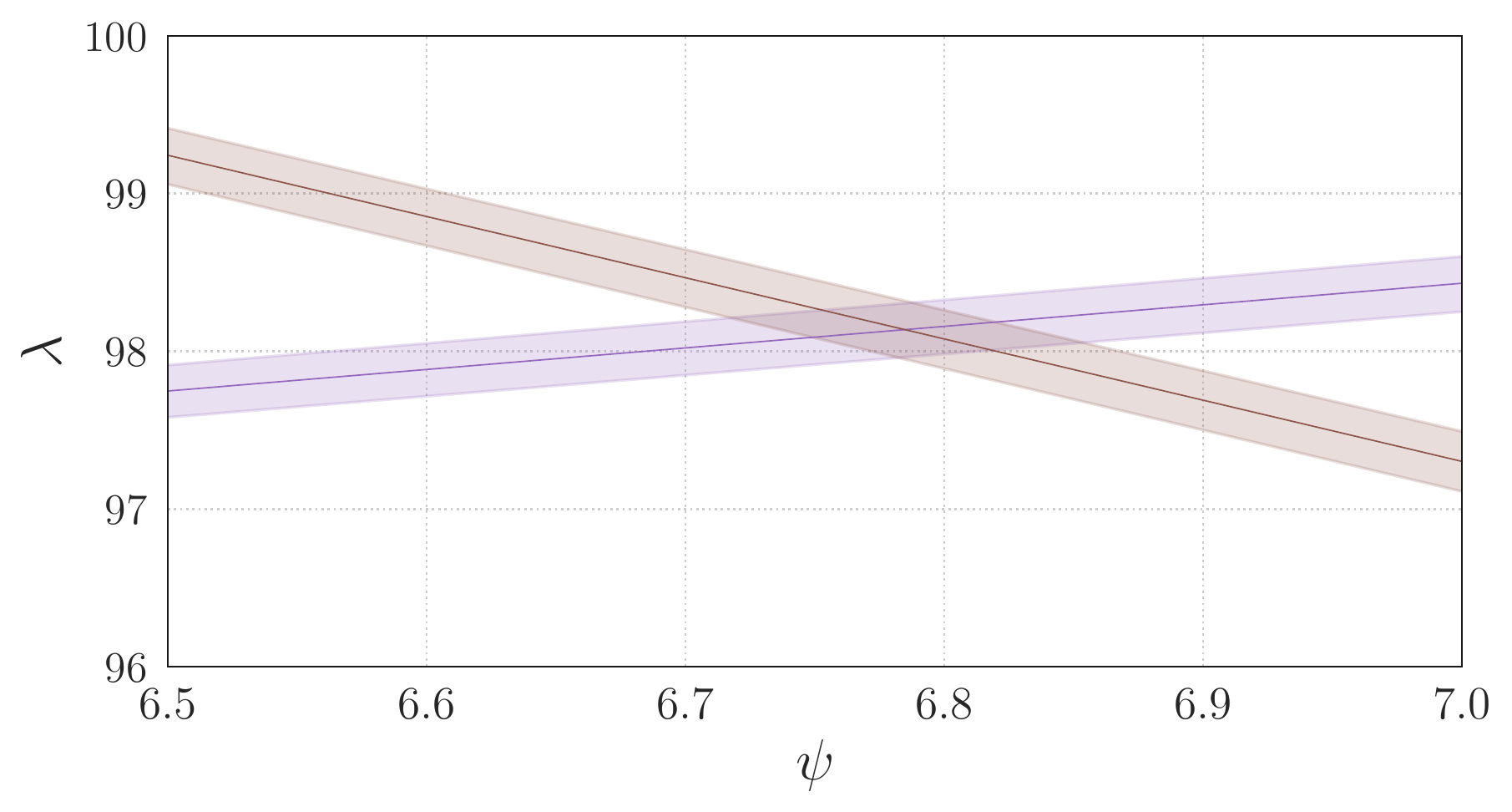}\quad&
\includegraphics[width=.31\textwidth]{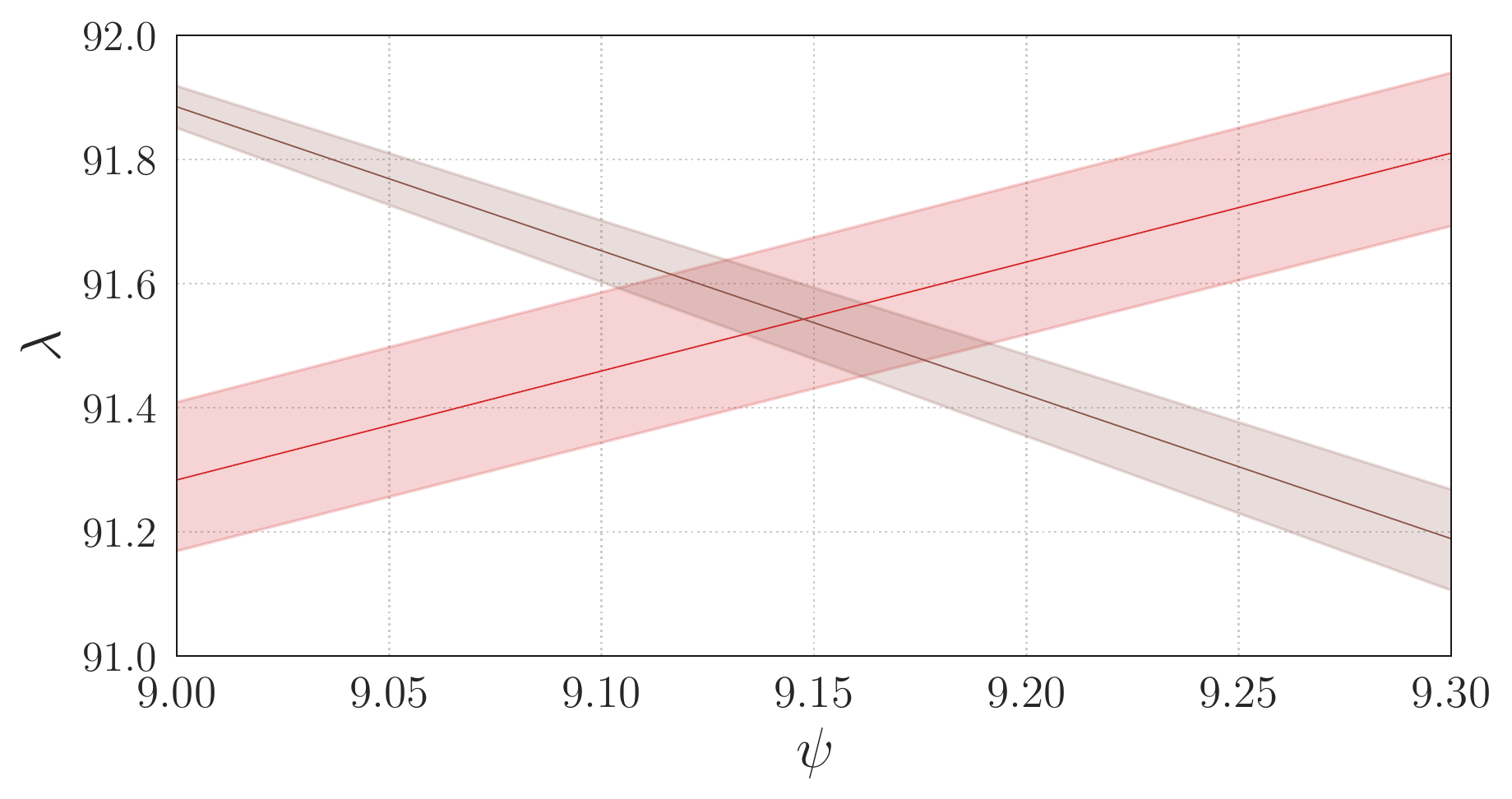}\\
\includegraphics[width=.28\textwidth]{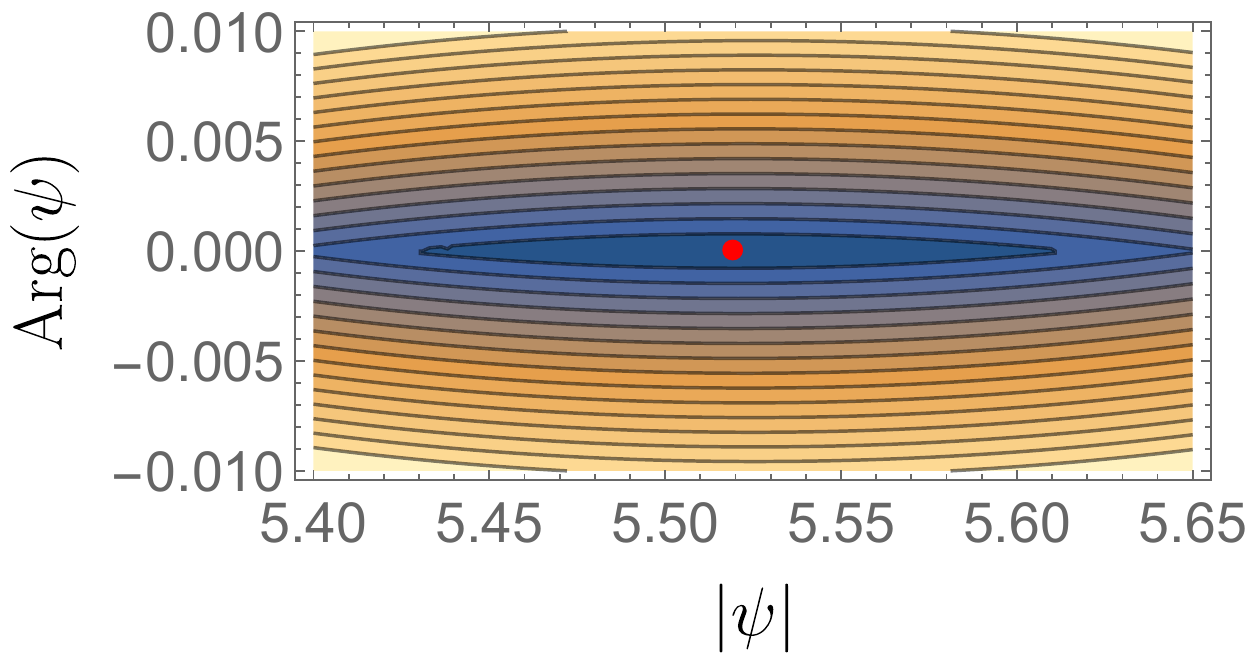}\quad&
\includegraphics[width=.28\textwidth]{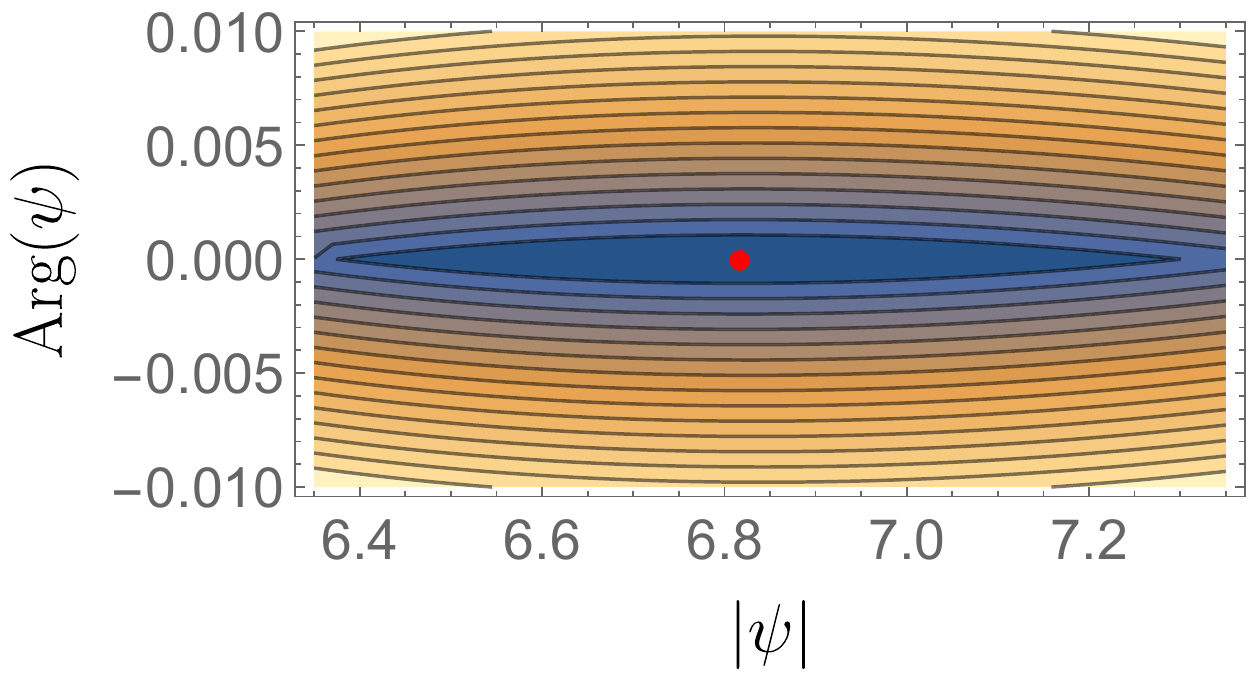}\quad&
\includegraphics[width=.28\textwidth]{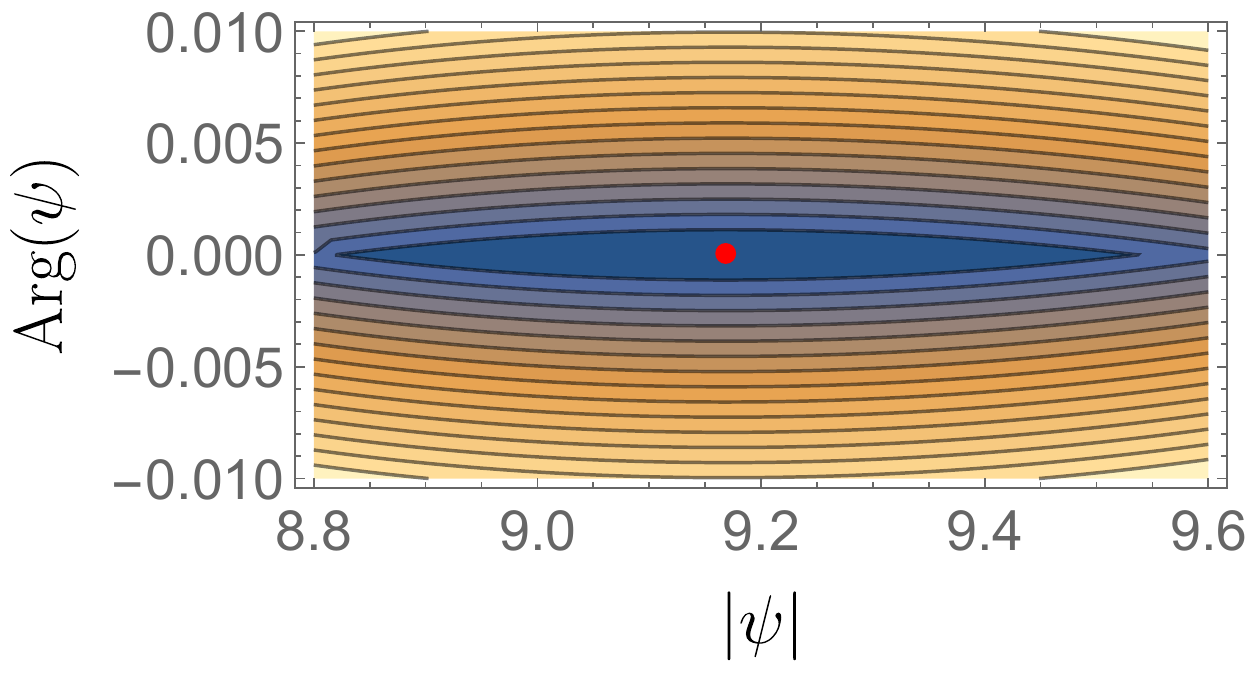}
\end{tabular}
\caption{A zoomed in view of the crossings observed in~\cite{Ashmore:2021qdf} for the one-parameter quintic along a slice of the moduli space with $\text{Im}(\psi)=0$. We find attractor points in the vicinity of the crossings, more precisely at $\psi=5.52$, $\psi=6.82$, and $\psi=9.17$, respectively.}
\label{fig:QuinticCrossings}
\end{figure}

Instead of solving the attractor flow equation~\eqref{eq:AttractorFlow}, we proceed by reading off the values of $\psi$ where eigenmodes cross and then compute the central charge~\eqref{eq:CentralCharge} for this choice of $\psi$ and a set of D-brane charges and look for a minimum of $|Z|$ in the vicinity of $\psi$. In practice, we do that by brute-force  scanning over D-brane charges and minimizing $|Z|$ numerically with the $\psi$ we read off from the crossings as a starting point. For the value of the crossings, we simply take some of the ones observed in~\cite{Ashmore:2021qdf}, which appear at around
\begin{align}
\label{eq:CrossingPointsQuintic}
\psi_1\approx 5.4\,,\qquad \psi_2\approx6.8\,,\qquad \psi_2\approx9.1\,,
\end{align}
cf.~Figure~\ref{fig:QuinticCrossings}.

When computing the central charge for these crossings, we should note that they were taken along a trajectory with $\text{Im}(\psi)=0$. Since $\psi\sim e^{2\pi i k/5}\psi$ for $k\in\mathbbm{Z}$, this means that we need to apply a monodromy matrix around the MUM point and shift the argument $\text{Arg}(\psi)\to2\pi/5 - \text{Arg}(\psi)$ upon computing the central charge in the vicinity of a crossing point with $\text{Arg}(\psi)<0$.
We present the results in Figure~\ref{fig:QuinticCrossings}. For each crossing value, we can identify a set of D6-D4-D2 or D4-D2-D0 charges that has an attract very close to the crossing points~\eqref{eq:CrossingPointsQuintic}.

%%%%%%%%%%%%%%%%%%%%%%%%%%%%%%%%%%%%%%%%%%%%%%%%%%%%%%%%%%%%%%%%%%%%%%%%%%%%
\section{Conclusion and Outlook}
\label{sec:Conclusions}
In this paper, we study the spectrum of the scalar Laplacian on a one-parameter family of CY $n$-folds for $n\leq3$. Our motivation comes explaining explain the behavior of the eigenmodes of the Laplace operator on the quintic observed in~\cite{Ashmore:2021qdf}: some eigenvalues become heavier and others lighter as we vary the complex structure, such that eigenmodes cross along special loci in complex structure moduli space. To develop intuition for the phenomenon, we first analyze CY one-folds, for which we have full analytic control. Given that the Ricci-flat CY metric is just flat, it is easy to write down a set of eigenfunctions and their eigenvalues, which are given in terms of winding numbers around the two periods of the torus. These are obtained from the underlying Picard-Fuchs system as a function of the complex structure modulus $\psi$ that appears in the defining equation of the torus described as a hypersurface in $\mathbbm{P}^2$, or by mapping the torus to Weierstrass form and reading of the Eisenstein series and the Klein $j(\tau)$ function to obtain the modular parameter $\tau$ in terms of $\psi$.

The analytic eigenvalues depend inversely on the area of the torus (which is given by the imaginary part $\tau_2$ of $\tau$) and on the winding numbers around the two cycles. As a consequence, eigenmodes that only wrap cycle of length $1$ become lighter as $\tau_2$ gets larger due to the 1/area factor, while for generic eigenmodes the factor of $\tau$ in the numerator competes with the area suppression in the denominator, leading to those eigenmodes either becoming heavier, staying constant, or becoming lighter. A similar mechanism could be at work for the quintic: the geodesic studied in~\cite{Ashmore:2021qdf} along which the behavior was observed started close to the conifold point and ran towards the MUM point at infinite distance. Close to the conifold point, the manifold degenerates, and a three-cycle (an $S^3$) shrinks, while close to the infinite complex structure point another three-cycle (a $T^3$) shrinks. This behavior is not special to the quintic under consideration but true in general if the SYZ conjecture is true (although in the case at hand, the $S^3$ and $T^3$ have been constructed explicitly by Candelas et.al.). Thus, there are also two competing cycles that can cause a similar behavior.

In general, crossings of two eigenmodes of the family of CY manifolds studied here appear along a real codimension 1 line inside their real two-dimensional complex structure moduli space. For the torus, we point out that along a real codimension 1 slice (along which the crossings appear at points), crossings are at Complex Multiplication points. In this sense, points where eigenvalues cross are special, even though nothing special seems to happen from the point of view of the defining equation or the moduli space metric.

For generalizing this result, we need to compute crossing points for higher $n$-folds, which requires resorting to numerical techniques. To get an understanding of the influence that different approximations can have on the eigenvalues, we compute them first numerically and vary the different hyperparameters entering the computation, and compare them to the exact, analytic results for the torus. These hyperparameters include the number of discrete points which we sample to approximate the CY, the quality of approximation of the Ricci-flat metric, how close we are to a degeneration point in complex structure moduli space, and the order at which we truncate the eigenbasis for the Laplacian eigenfunctions. We find that in general the metric dependence is not very pronounced, and  good results are already obtained from simply pulling back the ambient space Fubini-Study (FS) metric. This is interesting because obtaining an approximation to the Ricci-flat metric is computationally costly, even when using modern neural network techniques, while computing the FS metric is cheap. When truncating the eigenbasis to some number $k$ basis functions pulled back from the ambient space, we find that the lowest $2k/3$ eigenmodes can be approximated well, while for the last 33\% the finite truncation affects the accuracy. Sampling more points drastically improves the accuracy in the beginning but has essentially no effect once a threshold number of points is reached (10k for the torus). Finally, we observed that the sampling method is not sampling uniformly from the fundamental domain of the torus; horizontal voids develop that are due to how the hypersurface is embedded in $\mathbbm{P}^2$ and that become more pronounced as we move towards the infinite complex structure point. While the distribution of points obtained from the sampling method is known and we can just weight points in the over/undersampled regions differently, these voids still introduce numerical inaccuracies which get more and more pronounced. Knowing the dependence of the spectrum on these parameters is helpful for the rest of the paper, but also of independent interest for other applications involving the Laplacian eigenmodes.

Armed with an understanding of the error in the spectrum computation, we turn to a generalization of the notion of CM points o arbitrary CY $n$-folds, where it is phrased as a condition on the middle cohomology lattice. For $T^2$ and $K3$, CM points are closely related to attractor points in complex structure moduli space; however, in general the two are different. We argue that if the result from the torus generalizes, it is more likely that crossing points correspond to (rank 1) attractor points rather than CM points, since the latter are scarce, while the number of crossing points is infinite. We check this relation by computing the eigenspectrum numerically and reading off crossing points, which we then compare to known CM points for K3. For the quintic, we solve the attractor equation and show that there exists a D-brane system that leads to attractor points compatible with the crossing points. This is surprising since attractor points are related to BPS objects, while we are not aware of such a relation for higher eigenmodes of the Laplace operator. We hope to study this further in the future and, if correct, establish this connection.

\section*{Acknowledgments}
We thank Anthony Ashmore, Sergei Gukov, Sarah Harrison, Andre Lukas and Paul K.~Oehlmann for useful discussions. The work of FR is supported by the NSF grants PHY-2210333 and PHY-2019786 (The NSF AI Institute for Artificial Intelligence and Fundamental Interactions). The work of HA and FR is also supported by startup funding from Northeastern University.
%%%%%%%%%%%%%%%%%%%%%%%%%%%%%%%%%%%%%%%%%%%%%%%%%%%%%%%%%%%%%%%%%%%%%%%%%%%%
\bibliographystyle{bibstyle}
\bibliography{refs}
\end{document}